\documentclass[
superscriptaddress,nofootinbib,amsmath,amssymb,aps,longbibliography]{revtex4-2}
\usepackage[colorlinks=true,linkcolor=blue,citecolor=magenta,linktocpage=true]{hyperref}
\usepackage{graphicx}
\usepackage{hyperref}
\usepackage{mathtools}
\usepackage{comment}
\usepackage{url}
\usepackage{slashed}

\newcommand{\mrm}[1]{_{\rm #1}}

\renewcommand{\d}{{\rm d}}
\newcommand{\citecode}[1]{\vspace{0.1cm}\noindent\texttt{>>> #1}\vspace{0.1cm}}
\newcommand{\meanN}{\langle N\rangle}

\begin{document}

\thispagestyle{empty}

\begin{flushright}

CERN-TH-2019-067

\end{flushright}

\vspace{1.cm}

\begin{center}
 \large\textbf{BlackHawk v2.0: A public code for calculating the Hawking evaporation spectra of any black hole distribution}
\end{center}

\vspace{1.cm}

\begin{center}
	Alexandre Arbey$^{a,b,c,}$\footnote{\href{mailto:alexandre.arbey@ens-lyon.fr}{\texttt{alexandre.arbey@ens-lyon.fr}}} and J\'er\'emy Auffinger$^{a,}$\footnote{\href{mailto:j.auffinger@ipnl.in2p3.fr}{\texttt{j.auffinger@ipnl.in2p3.fr}}}\\[0.4cm]
	{\sl$^a$Univ Lyon, Univ Claude Bernard Lyon 1,\\ CNRS/IN2P3, IP2I Lyon, UMR 5822, F-69622, Villeurbanne, France}\\
	\vspace{0.2cm}
	{\sl$^b$Theoretical Physics Department, CERN, CH-1211 Geneva 23, Switzerland}\\
	\vspace{0.2cm}
	{\sl$^c$Institut Universitaire de France (IUF), 103 boulevard Saint-Michel, 75005 Paris, France}
\end{center}

\vspace{1.cm}

{\bf URL:} \url{http://blackhawk.hepforge.org/}

\vspace{0.5cm}

{\bf arXiv identifier:} \href{http://arxiv.org/abs/1905.04268}{1905.04268 [gr-qc]}

\vspace{1.cm}

\begin{center}
{\bf Abstract}
\end{center}
We describe \texttt{BlackHawk}, a public C program for calculating the Hawking evaporation spectra of any black hole distribution. This program allows the users to compute the primary and secondary spectra of stable or long-lived particles generated by Hawking radiation of the distribution of black holes, and to study their evolution in time. The physics of Hawking radiation is presented, and the capabilities, features and usage of \texttt{BlackHawk} are described here under the form of a manual. This is the \texttt{BlackHawk v2.0} manual, which is available on the \texttt{BlackHawk} webpage \url{http://blackhawk.hepforge.org/}. A brief release note summarizing the new aspects of \texttt{BlackHawk v2.0} as well as illustrating examples can be found in~\cite{release_note}.

\newpage

\tableofcontents{}

\newpage



\section{Introduction}
\label{sec:introduction}

Black holes (BHs) may be the most promising astrophysical objects in the study of beyond general relativity (GR) and beyond Standard Model (SM) physics. BHs of stellar mass $M\sim 5-100\,M_\odot$ have been observed indirectly from X-ray emission of binaries for quite some time (see \textit{e.g.}~\cite{Motta:2021zlt} and references therein). More recently, gravitational wave (GW) instruments such as LIGO/VIRGO have detected the perturbation of spacetime generated by the merging of stellar mass BHs, confirming GR predictions with great accuracy~\cite{LIGOScientific:2018mvr,LIGOScientific:2020tif,LIGOScientific:2020kqk,LIGOScientific:2020ibl}. Future instruments may improve over these catalogs, giving access to relevant BH population statistics and to an extended mass and binary separation ranges with unprecedented precision (these include KAGRA~\cite{KAGRA:2020cvd}, the Einstein Telescope~\cite{Punturo:2010zz} and LISA~\cite{LISA:2017pwj}). Supermassive BHs with $M\gg 10^{6}\,M_\odot$ are believed to lie at the centre of every big galaxy and have been detected by radio interferometers; such as the central BH of M87 (see~\cite{EventHorizonTelescope:2019dse}, and subsequent papers for details). There are no observation in the intermediate mass range yet, but GW interferometers could detect $M\sim 10^{2}-10^{3}\,M_\odot$ in the near future~\cite{LIGOScientific:2021tfm}. Interestingly, the first BH population statistics extracted from these catalogs are not easily explained by stellar evolution. Thus, some claim that part of the observed stellar mass BHs are in fact of primordial origin, that is to say BHs formed in the early universe through the collapse of overdensities (for recent reviews on primordial BHs --- PBHs --- see \textit{e.g.}~\cite{Carr:2020gox} and~\cite{Green:2020jor}). This channel of formation seems to be consistent with the LIGO/VIRGO observations if dark matter (DM) is composed at most of $\sim1\%$ of stellar mass PBHs~\cite{Carr:2020gox,DeLuca:2021wjr}. Future observations such as the precise measure of the spin of the binary BHs components~\cite{Arbey:2019jmj}, the stochastic GW background~\cite{Mukherjee:2021ags} or detection of BHs with masses outside the stellar evolution predictions~\cite{Phukon:2021cus,LIGOScientific:2020iuh} could point toward a primordial origin for some of those; while it is still not clear how supermassive BHs can be present at such high redshifts: they may have been formed from the hierarchical mergers of massive PBH seeds~\cite{Carr:2018rid}. On the other extreme of the mass range, PBHs could be as small as the Planck mass $M\sim 10^{-5}\,$g, even if some inflation models place lower bounds of $M\gtrsim 10^{-1}\,$g~\cite{Planck:2018jri,Carr:2020gox}. Their detection represents a challenge of modern astrophysics (see the review~\cite{Carr:2020gox}).

Hawking proved that BHs emit radiation as blackbodies, therefore slowly losing mass during a process called Hawking evaporation~\cite{Hawking:1975vcx}. One major feature of that process is that Hawking radiation (HR) gets more and more energetic as the BH gets lighter, reaching the Planck energy at the end of the BH evaporation. Of course, BHs with initial mass $M\lesssim M_\odot$ (below the Oppenheimer--Volkoff limit) are presumably of primordial origin, but even stellar mass BHs would evaporate away within an incredibly long time. Tiny PBHs of $\sim 10^{15}-10^{18}\,$g emit gamma rays that could be detected in Earth of space based observatories. PBHs even smaller, with initial mass $M\lesssim 10^{15}\,$g, would have already disappeared and can not represent a fraction of DM today. Their evaporation in the early universe could have left imprints in cosmological observables~\cite{Carr:2020gox}. Larger PBHs with $M\gtrsim 10^{18}\,$g still emit HR but with a power that is so small that detection is considered out of reach for now (see however the proposition~\cite{Arbey:2020urq}). Historically, Hawking radiation has been considered as a mean of detection of tiny BHs evaporating instantaneously in high energy colliders such as the LHC, within models where extra spatial dimensions reduce the Planck mass and thus the energy density required to form a BH. The public codes \texttt{BlackMax}~\cite{Dai:2009by} and \texttt{Charybdis}~\cite{Frost:2009cf} were published in that context: they compute the spectrum of particles generated by the evaporation of higher-dimensional (spinning) BHs in a LHC-like detector. As HR is a purely gravitational phenomenon, all particles of the (beyond the) SM spectrum are emitted. These particles are subsequently subject to (beyond the) SM interactions such as hadronization, inter-conversion and decay, before reaching some detector. Hence, particle physics codes such as \texttt{PYTHIA}~\cite{Sjostrand:2014zea}, \texttt{HERWIG}~\cite{Bellm:2015jjp} or \texttt{Hazma}~\cite{Coogan:2019qpu} have been used to compute the secondary spectra of particles resulting from this primary emission. There is therefore a strong link between HR and particle physics, particularly in the QCD domain~\cite{MacGibbon:1990zk,MacGibbon:1991tj}. However, before \texttt{BlackHawk} was released, there was no publicly available code to compute the Hawking radiation spectrum of a distribution of BHs in the universe, \textit{e.g.}~to derive precise constraints on the DM fraction under the form of PBHs or to predict detection signals of HR in future instruments; people usually relied on analytical approximations to compute the spectra. This was the motivation that drove the publication of the code. Moreover, since its first release, \texttt{BlackHawk} has received many modifications to embed interesting features: distributions of spinning BHs, emission of DM on top of the SM particles, HR rates in modified BH solutions, \textit{etc}. Some of these modifications were already present in the previous version of the manual on arXiv [\href{https://arxiv.org/abs/1905.04268v2}{1905.04268v2}], but most of them have been added in the recent update presented in the \texttt{BlackHawk v2.0} release note~\cite{release_note}. The latter also contains a complete overview of the publications realised since \texttt{BlackHawk v1.0} was released. The present manual is associated with \texttt{BlackHawk v2.0}, which is publicly available on the webpage:
\begin{center}
	\url{http://blackhawk.hepforge.org/}
\end{center}
It is structured as follows: in Section~\ref{sec:hawking_radiation} we review the physics and main formulas of Hawking radiation, in Section~\ref{sec:content_compilation} we provide the necessary compilation information, Section~\ref{sec:input_parameters} describes the parameters of a typical run of the code, Section~\ref{sec:routines} details the different routines, the two main \texttt{BlackHawk} programs are presented in Section~\ref{sec:programs}, the output files are explained in Section~\ref{sec:output_files}, we briefly estimate the memory usage of a run in Section~\ref{sec:memory_usage}, we outline some possible applications of the code in Section~\ref{sec:other_applications} and we conclude in Section~\ref{sec:conclusion}.


\section{Hawking radiation}
\label{sec:hawking_radiation}

In this Section we review the physics of Hawking radiation. We present the formulas and results in their most general forms, and use the natural system of units $G = \hbar = k\mrm{B} = c = 4\pi\varepsilon_0 = 1$ unless stated otherwise.

\subsection{Black hole distributions}
\label{subsec:black_hole_distributions}

\texttt{BlackHawk} has been designed to provide tests of compatibility between observations
and BH distributions at different main steps of the history of the universe. For
this purpose, it computes the Hawking emission of a distribution of BHs, and its
evolution in time. The obtained spectra can then be used to check whether the
amount of produced particles has an effect on observable cosmological quantities, such as the cosmic microwave background (CMB), extragalactic gamma-ray background (EGRB), the Big-Bang nucleosynthesis (BBN) yields, \textit{etc}.

The distribution of BHs as a function of their mass or any other parameter (spin, charge) is completely model dependent. In \texttt{BlackHawk}, we are agnostic about the channel of formation of BHs and thus provide the possibility to use any distribution. For primordial BHs formed in the radiation era after inflation, there is typically a general relationship between the PBH formation time $t\mrm{f}$ and its initial mass~\cite{Carr:2020gox,Auffinger:2020afu}
\begin{equation}
	t\mrm{f} \sim \dfrac{M}{\gamma}\,, \label{eq:formation_time}
\end{equation}
where $\gamma \simeq 0.2$ encodes the details of the gravitational collapse of an overdense region. Of course, before Hawking radiation becomes relevant and depending on the plasma environment of PBHs, accretion (or mergers) could change this initial time-mass relationship~\cite{Masina:2020xhk}. Usually, a monochromatic (Dirac delta) mass/spin function is used to constrain the abundance of BHs. However, recent studies have shown that the generalisation to \textit{e.g.}~extended mass functions is not trivial~\cite{Carr:2017jsz}. The cleanest way consists in recomputing the constraints taking into account the extended distribution, rather than extrapolating from the monochromatic results. \texttt{BlackHawk} can in principle work with any distribution of BHs. Several BH mass and secondary BH parameter distributions are already built-in and listed in the following. It is also possible that the user provides its own distribution for BH masses and their secondary parameters thanks to a \texttt{.txt} file.

\subsubsection{Mass distributions}

Here we list the \texttt{BlackHawk} built-in mass distributions $\d n/\d M$ such that the integral
\begin{equation}
	n\mrm{BH} = \int_{M\mrm{min}}^{M\mrm{max}}\dfrac{\d n}{\d M} \d M\,,
\end{equation}
gives the total number of comoving BHs per unit volume. To obtain the density of those BHs, one simply considers the integral
\begin{equation}
	\rho\mrm{BH} = \int_{M\mrm{min}}^{M\mrm{max}}M\dfrac{\d n}{\d M} \d M\,.
\end{equation}
In the following formulas, the normalization factor $A$ which appears for each distribution has a mass unit that depends on the details of the expression of $\d n/\d M$.

\paragraph{\textbf{Peak theory distribution}}

The peak theory distribution is derived from the scale-invariant model, assuming that the power spectrum of the primordial density fluctuations is a power-law (see \textit{e.g.}~\cite{Tashiro:2008sf,Germani:2018jgr})
\begin{equation}
	P(k) = R\mrm{c} \left( \frac{k}{k_0} \right)^{n-1},
\end{equation}
where the spectral index is $n\approx 1.3$ and $R\mrm{c}$ is measured using the CMB to be $R\mrm{c} = (24.0\pm 1.2)\times 10^{-10}$ at the scale $k_0 = 0.002\,$Mpc$^{-1}$. The comoving number density of BHs resulting from this power spectrum is obtained in~\cite{Tashiro:2008sf} through peak-theory
\begin{equation}
	{\rm d}n \approx \frac{1}{4\pi^2 M}\left( \frac{X(n-1)}{6M} \right)^{3/2}\frac{(n-1)}{2}\nu^4e^{-\nu^2/2}{\rm d}M\,, \label{eq:peak_theory}
\end{equation}
where
\begin{equation}
	\nu(M) \equiv \left( \frac{2(k_0^2 M/X)^{(n-1)/2}}{R\mrm{c}\Gamma((n-1)/2)} \right)^{1/2}\zeta\mrm{th}\,, \qquad X \equiv \frac{4\pi}{3}\left( \frac{8\pi G}{3} \right)^{-1}\left\{ \frac{H_0^2\Omega\mrm{m}}{1+z\mrm{eq}}\left( \frac{g_{*,\rm eq}}{g_*} \right)^{1/3} \right\}^{1/2}, \label{eq:nu}
\end{equation}
in which:
\begin{itemize}
	\item $H_0 = 67.8\,$km$\cdot$s$^{-1}\cdot$Mpc$^{-1}$ is the Hubble parameter~\cite{ParticleDataGroup:2018ovx},
	\item $\Omega\mrm{m} = 0.308$ is the matter energy density in the universe~\cite{ParticleDataGroup:2018ovx},
	\item $z\mrm{eq} = 3200$ is the radiation-matter equality redshift~\cite{Tashiro:2008sf},
	\item $g_{*,\rm eq} = 3.36$ is the number of relativistic energy degrees of freedom (dof) at radiation-matter equality~\cite{Tashiro:2008sf},
	\item $g_* = 106.75$ is the number of (SM) relativistic energy dof at the time of BH formation, here the end of the inflation~\cite{ParticleDataGroup:2018ovx},
	\item $\zeta\mrm{th} = 0.7$ is the overdensity threshold that provokes the direct collapse of a density fluctuation into a BH~\cite{Tashiro:2008sf}.
\end{itemize}

\paragraph{\textbf{Log-normal distributions}}

The log-normal distribution~\cite{Carr:2017jsz} is considered to be the general mass function originating from a peak in the power spectrum of primordial fluctuations. It is parametrized through
\begin{equation}
	{\rm d}n = \dfrac{ A}{\sqrt{2\pi}\sigma M^2}{\rm exp}\left( -\dfrac{\ln(M/M\mrm{c})^2}{2\sigma^2} \right){\rm d}M\,, \label{eq:lognormal}
\end{equation}
where $A$ is the amplitude, $M\mrm{c}$ is the position of the peak and $\sigma$ is its width. Note that this is a log-normal distribution for the comoving \textit{density} $M\d n/\d M$ and not for the comoving \textit{number density} $\d n/\d M$ -- which differs only by a factor of $M$ and a different normalization $A$
\begin{equation}
	{\rm d}n = \dfrac{ A}{\sqrt{2\pi}\sigma M}{\rm exp}\left( -\dfrac{\ln(M/M\mrm{c})^2}{2\sigma^2} \right){\rm d}M\,. \label{eq:lognormal2}
\end{equation}
These distributions reduce to the monochromatic distribution in the limit $\sigma\rightarrow 0$, which can be used as a consistency check.

\paragraph{\textbf{Power-law distribution}}

The power-law distribution~\cite{Carr:2017jsz} is a less refined version of Eq.~\eqref{eq:peak_theory}. It also derives from scale-invariant primordial density fluctuations and is given by
\begin{equation}
	{\rm d}n = AM^{\gamma-2}{\rm d}M\,, \label{eq:powerlaw}
\end{equation}
where $\gamma \equiv -2w/(1+w)$ and $w$ is defined through the equation of state of the energy density that dominates the universe at the epoch of BH formation such that $P=w\rho$, see Table~\ref{tab:eq_state} below.

\begin{table}[!ht]
	\centering{
		\begin{tabular}{|c|c|}
			\hline
			Fluid & Equation of state \\
			\hline
			Matter & $w = 0$ \\
			Radiation & $w = 1/3$ \\
			Cosmological constant & $w = -1$ \\
			Curvature & $w = -1/3$ \\
			\hline
		\end{tabular}
		\caption{Equation of state for different cosmological fluids.\label{tab:eq_state}}
	}
\end{table}

\paragraph{\textbf{Critical collapse distribution}}

The critical collapse distribution~\cite{Carr:2017jsz} derives from a Dirac power spectrum for the primordial density fluctuations. It is defined as
\begin{equation}
	{\rm d}n = AM^{1.85}\exp\left( -\left(\dfrac{M}{M\mrm{f}}\right)^{2.85} \right){\rm d}M\,, \label{eq:criticalcollapse}
\end{equation}
where $A$ is an amplitude factor and $M\mrm{f}$ an upper cut-off.

\paragraph{\textbf{Uniform distribution}}

The uniform distribution is a toy-model defined as
\begin{equation}
	{\rm d}n = \dfrac{A}{M\mrm{max} - M\mrm{min}}{\rm d}M\,, \label{eq:uniform_M}
\end{equation}
where $A$ is an amplitude factor and $M\mrm{min,max}$ are the distribution bounds.

\paragraph{\textbf{Dirac distribution}}

The Dirac distribution simulates a single-valued BH mass function $M = M\mrm{c}$
\begin{equation}
	\d n = A\delta(M - M\mrm{c})\,.
\end{equation}
It is useful to perform monochromatic analyses and consistency checks for a single BH. It is by default normalized to 1 BH per comoving cm$^3$, any additional amplitude factor $A$ must be applied afterwards.

\subsubsection{Secondary parameter distribution}

Here we list the secondary BH parameter distributions. The no-hair theorem states that in general relativity, BHs are exclusively described by the set $\{M,a^*,Q^*\}$ where $a^* \equiv a/M \equiv J/M^2$ is the adimensioned BH angular momentum and $Q^* \equiv Q/M$ is the adimensioned BH electric charge. The three parameters are constrained by
\begin{equation}
	a^{*2} + Q^{*2} < 1\,,
\end{equation}
otherwise the interior Cauchy horizon would collapse with the BH horizon and the singularity would be exposed, violating the GR singularity conjecture. Therefore, in GR we have two possible secondary BH parameters, namely the charge $Q^*$ and the angular momentum $a^*$. As is well known, GR can not be the fundamental description of gravity, as it is unable to provide a quantum mechanics framework for the latter. Lots of alternative models have been proposed, most of which include additional secondary parameters, \textit{e.g.} the number of extra dimensions $n$ in higher-dimensional theories of gravity or the polymerization parameter $\varepsilon$ in polymerization models deriving from loop quantum gravity. Obviously, the two last examples of secondary parameters are constants common to all the BHs of a given model --- the only physical distribution of those is a Dirac function. However, realistic models of BH formation should include an extended distribution of both angular momentum and electric charge, whose functional form depends on the BH formation mechanism. For example, most models predict that BHs form in electrically neutral environments, either from the collapse of stars or in the primordial universe. A phase transition in the early universe could however produce regions of non-zero electric charge that collapse to form charged PBHs. On the other hand, there are numerous models describing the spin distribution of BHs~\cite{Flores:2021tmc}, which could originate from accretion~\cite{LKrol:2021pau}, hierarchical mergers~\cite{Fishbach:2017dwv,Doctor:2021qfn}, close encounters in BHs clusters~\cite{Nelson:2019czq,Jaraba:2021ces} or the initial angular momentum or non-sphericity of the collapsing region~\cite{Harada:2016mhb}. Interestingly, the spin of BHs could act as a distinguishing rule between primordial or stellar origin (\textit{e.g.}~\cite{Arbey:2019jmj,DeLuca:2020bjf}), or increase the detection possibilities (\textit{e.g.}~\cite{Arbey:2019vqx,Kuhnel:2019zbc,Arbey:2021ysg}). Inside \texttt{BlackHawk}, we have built some benchmark distributions $\d \tilde{n}/\d x$ for both the angular momentum and the charge $x = a^*,Q^*$, with an integral normalized to unity
\begin{equation}
	\tilde{n}\mrm{BH} = \int_0^1 \dfrac{\d \tilde{n}}{\d x}\d x = 1\,.
\end{equation}

\paragraph{\textbf{Gaussian distribution}}

The Gaussian distribution is a generalisation of the monochromatic distribution, which reduces to the latter in the limit $\sigma \rightarrow 0$
\begin{equation}
	{\rm d}\tilde{n} = \dfrac{1}{\sqrt{2\pi\sigma^2}}\exp\left( -\dfrac{(x - x\mrm{c})^2}{2\sigma^2} \right){\rm d}x\,,\label{eq:gaussian_param}
\end{equation}
where $\sigma$ is the standard deviation and $x\mrm{c}$ a characteristic value of $x$.

\paragraph{\textbf{Uniform distribution}}

This is the same distribution as for the mass, but normalized to unity
\begin{equation}
	{\rm d}\tilde{n} = \dfrac{{\rm d}x}{x\mrm{max} - x\mrm{min}}\,,\label{eq:uniform_param}
\end{equation}
where $x\mrm{min,max}$ are the distribution bounds.

\paragraph{\textbf{Dirac distribution}}

This is the same distribution as for the mass, and it also applies for all the possible single-valued secondary parameters $x = x\mrm{c}$
\begin{equation}
	\d \tilde{n} = \delta(x - x\mrm{c})\,.
\end{equation}

\subsection{Hawking evaporation physics}

In this Section we review the basic formulas for the Hawking evaporation process. Hawking radiation is a semi-classical phenomenon arising from the mix between classical general relativity and aspects of quantum mechanics. Random quantum fluctuations at the horizon of BHs create pairs of particles and anti-particles. It is then possible that (anti-)particles escape to spatial infinity, giving rise to a net flux of outgoing radiation denoted as Hawking radiation. This process is qualitatively and quantitatively linked to the details of the BH solution metric; there are many such solutions of the Einstein equations of general relativity, not all of which exhibit Hawking radiation. For the spherically symmetric and static BHs, we rely on the two companion papers~\cite{Arbey:2021jif,Arbey:2021yke} (see relevant extended bibliography there).

\subsubsection{Spherically symmetric and static black holes}

The general metric of a spherically symmetric and static BH solution is in Boyer--Lindquist coordinates
\begin{equation}
	\d s^2 = -G(r)\d t^2 + \dfrac{1}{F(r)}\d r^2 + H(r)\d \Omega^2\,,
\end{equation}
where $\d\Omega^2 \equiv \d\theta^2 + \sin(\theta)\d\varphi^2$ is the 4-dimensional 2-sphere element. We also require that this solution is asymptotically flat, meaning that
\begin{equation}
	F(r),G(r) \underset{r\rightarrow +\infty}{\longrightarrow} 1\,, \qquad H(r) \underset{r\rightarrow+\infty}{\sim} r^2\,,
\end{equation}
and that $F(r)$ exhibits at least one pole at $r = r\mrm{H}$, the BH horizon. Examples used in \texttt{BlackHawk} are listed below.

\paragraph{\textbf{Schwarzschild black hole}} This is the simplest example of a BH solution, describing the spacetime curvature around a non-rotating, uncharged mass $M$ enclosed by a horizon at $r\mrm{H} = 2M \equiv r\mrm{S}$. The metric coefficients are
\begin{equation}
	F(r) = G(r) = 1 - \dfrac{r\mrm{H}}{r}\,, \qquad H(r) = r^2\,.
\end{equation}
It is an example of the common $tr$-symmetric metrics with $F(r) = G(r)$ and $H(r) = r^2$. All other BH solutions embed this case as a limiting behaviour.

\paragraph{\textbf{Charged black hole}} Reissner--Nordstr\"om BHs describe the spacetime around a mass $M$ with charge $Q^* \equiv Q/M$. The $tr$-symmetric metric coefficients are
\begin{equation}
	F(r) = G(r) = 1-\dfrac{r\mrm{S}}{r} + \dfrac{r\mrm{Q}^2}{r^2}\,, \qquad H(r) = r^2\,,
\end{equation}
where $r\mrm{Q}^2 = Q^2$ in our system of units. The Hawking horizon $r\mrm{H}$ and Cauchy horizon $r\mrm{C}$ are the poles of $F(r)$, respectively
\begin{equation}
	r\mrm{H,C} \equiv r_{\pm} \equiv r\mrm{S}\dfrac{1\pm\sqrt{1-4r\mrm{Q}^2/r\mrm{S}^2}}{2}\,.\label{eq:rH_charged}
\end{equation}
We recover the Schwarzschild solution in the limit $Q^*\rightarrow 0$ (in that case, $r\mrm{Q} = 0 = r_-$). The opposite limit $Q^*\rightarrow 1$ is called ``extremal" BH and $Q^* = 1$ is not physical.

\paragraph{\textbf{Higher-dimensional black hole}} Models of general relativity with higher dimensions (more than 3 spatial dimensions) are not excluded yet by particle physics experiments or cosmological observations if the additional spatial dimensions are small enough and compactified~\cite{Johnson:2020tiw}. The $tr$-symmetric metric coefficients are
\begin{equation}
	F(r) = G(r) = 1-\left(\dfrac{r\mrm{H}}{r}\right)^{n+1}, \qquad H(r) = r^2\,,
\end{equation}
where $n = 0, 1, \dots$ is the number of extra spatial dimensions and
\begin{equation}
	r\mrm{H} = \dfrac{1}{\sqrt{\pi}M_*}\left(\dfrac{M}{M_*}\right)^{1/(n+1)}\left(\dfrac{8\Gamma\big((n+3)/2\big)}{n+2}\right)^{1/(n+1)}, \label{eq:rH_higher}
\end{equation}
is the BH horizon with $\Gamma$ the Euler gamma function. In these theories, the Planck mass is rescaled by $M\mrm{P}^2 \sim M_*^{n+2} R^n$ where $R$ is the typical size of the extra dimensions. The 4-dimensional Schwarzschild case corresponds to $n = 0$.

\paragraph{\textbf{Polymerized black hole}} Polymerized BHs are inspired by the quantization procedure in loop quantum gravity. They are part of a more general class of BH solutions that solve the central singularity problem, as the metric coefficients
\begin{equation}
	F(r) = \dfrac{(r-r_+)(r-r_-)r^4}{(r+r_*)^2(r^4+a_0^2)}\,, \qquad G(r) = \dfrac{(r-r_+)(r-r_-)(r+r_*)^2}{r^4+a_0^2}\,, \qquad H(r) = r^2+\dfrac{a_0^2}{r^2}\,,
\end{equation}
do not diverge at $r\rightarrow 0$. The horizon and Cauchy radii are
\begin{equation}
	r\mrm{H} \equiv r_+ \equiv 2m(\varepsilon)\,, \qquad r\mrm{C} \equiv r_- \equiv 2m(\varepsilon)P(\varepsilon)^2\,,\label{eq:rH_polymerized}
\end{equation}
where $\varepsilon$ is the polymerization parameter which describes the scale of the quantum corrections, $P(\varepsilon)$ is the polymerization function
\begin{equation}
	P(\varepsilon) = \dfrac{\sqrt{1 + \varepsilon^2} - 1}{\sqrt{1 + \varepsilon^2} + 1}\,, \label{eq:poly_function}
\end{equation}
and the Arnowitt--Deser--Misner mass is given by
\begin{equation}
	M = m(\varepsilon)(1+P(\varepsilon))^2\,.\label{eq:m_LQG}
\end{equation}
The parameter $a_0$ is the area gap in loop quantum gravity. We recover the Schwarzschild case in the limits $\varepsilon,a_0 \rightarrow 0$.

\subsubsection{Rotating black holes}

Realistic BH formation channels predict a non-zero angular momentum. The metric of a (rotating) Kerr BH is given by
\begin{equation}
	\d s^2 = \big(\d t - a \sin^2(\theta) \d\phi\big)^2\, \frac{\Delta}{\Sigma} - \left(\frac{\d r^2}{\Delta} + \d\theta^2\right) \Sigma - \big((r^2+a^2)\d\phi-a \d t\big)^2 \, \frac{\sin^2(\theta)}{\Sigma}\,,
\end{equation}
where $\Sigma(r) \equiv r^2 + a^2 \cos^2(\theta)$ and $\Delta(r) \equiv r^2 - 2Mr + a^2$. The horizon and Cauchy radii are
\begin{equation}
	r\mrm{H,C} \equiv r_\pm \equiv r\mrm{S}\dfrac{1 \pm \sqrt{1 - a^{*2}}}{2}\,.\label{eq:rH_Kerr}
\end{equation}
It is not trivial at all to mix this BH solution with the (beyond GR) solutions listed above. Thus, in \texttt{BlackHawk}, we keep them separated and do not consider \textit{e.g.} Kerr--Newman rotating \textit{and} charged BHs. We recover the Schwarzschild case in the limit $a^* \rightarrow 0$, while the other limit $a^* \rightarrow 1$ is called ``extremal" (the case $a^* = 1$ is not physical).

\subsubsection{Particle creation by black holes}

\paragraph{\textbf{Black hole temperature}}

Once the metric of spacetime around a BH is known, Hawking radiation rates are computed as follows. First, the deformation of spacetime is associated with an effective horizon temperature
\begin{equation}
	T = \dfrac{\kappa}{2\pi}\,,
\end{equation}
where $\kappa$ is the surface gravity of the BH. In the case of a spherically symmetric and static metric, it is given by~\cite{Arbey:2021yke}
\begin{equation}
	\kappa = \left.\dfrac{1}{4}\frac{FG^{\prime 2}}{G}\right|_{\rm r = r\mrm{H}}.
\end{equation}
For a Kerr BH, it is
\begin{equation}
	\kappa = \dfrac{r_+ - M}{r_+^2 + a^2}\,.
\end{equation}
We can point out that the temperature of an ``extremal" BH with $a^* = 1$, $Q^* = 1$ or $\varepsilon \rightarrow +\infty$ is 0, which is thermodynamically not physical. The temperatures associated with the different metrics listed above are summarized in Table~\ref{tab:temperatures}.

\begin{table}[!ht]
	\centering{
		\begin{tabular}{|c|c|}
			\hline
			Metric & Temperature \\ \hline
			Schwarzschild & $T\mrm{S} = \dfrac{1}{4\pi r\mrm{S}}$ \\ \hline
			Reissner--Nordstr\"om & $T\mrm{Q} = \dfrac{r_+ - r_-}{4\pi r_+^2}$ \\ \hline 
			Higher-dimensional & $T_n = \dfrac{n+1}{4\pi r\mrm{H}}$ \\ \hline 
			Polymerized & $T\mrm{LQG} = \dfrac{r_+^2(r_+ - r_-)}{4\pi(r_+^4 + a_0^2)}$ \\ \hline
			Kerr & $T\mrm{K} = \dfrac{1}{2\pi}\left( \dfrac{r_+ - M}{r_+^2 + a^2} \right)$ \\
			\hline
		\end{tabular}
		\caption{Horizon temperatures for different BH metrics.\label{tab:temperatures}}
	}
\end{table}

\paragraph{\textbf{Emission rates}} Then, the emission rate of some (beyond the) SM degrees of freedom $i$ per unit time and energy is given by the master equation~\cite{Hawking:1975vcx}
\begin{equation}
	\dfrac{\d^2 N_{i,lm}}{\d t\d E} = \dfrac{1}{2\pi} \dfrac{\Gamma_{s_i l m}(E,M,x_j)}{e^{E^\prime/T}  - (-1)^{2s_i}}\,. \label{eq:hawking_master}
\end{equation}
In this equation, we identify the Boltzmann statistics factor with the BH temperature in the denominator, with $\pm 1$ for fermions and bosons respectively. The particle energy $E$ has to be corrected for couplings between the emitted particle and the BH; in \texttt{BlackHawk} the only correction is brought by Kerr BHs for which the horizon rotation gives
\begin{equation}
	E^\prime\mrm{K} \equiv E - m\Omega\,, \qquad \Omega \equiv \dfrac{a^*}{(2r_+)}\,.
\end{equation}
We do not consider inside \texttt{BlackHawk} the electric interaction between a Reissner--Nordstr\"om BH and a charged particle. In these equations, we have decomposed the wave function of the particle fields on the base of the spin-weighted spheroidal harmonics of spin $s_i$, and $l,m$ are respectively the particle angular momentum $l \ge s_i$ and its projection on the BH axis $m = -l,\dots,+l$. The last factor we need to determine is the greybody factor $\Gamma_{s_i l m}(E,M,x_j)$. This factor encodes the probability that a (anti-)particle generated at the BH horizon escapes to spatial infinity. It depends on the particle energy $E$, the BH mass $M$, and the set of secondary parameters for the BH metric $x_j$ (\textit{e.g.}~the BH spin $a^*$). It depends \textit{a priori} on the particle rest mass $\mu_i$, but for simplicity we apply the kinematic condition $E>\mu_i$ as a simple cut-off in the emission rates of massive particles. The computation of the greybody factors inside \texttt{BlackHawk} is detailed in Appendix~\ref{app:greybody_factors}. In order to obtain the emission rate of some particle $i$, we sum over the field angular momentum $l,m$, helicity $g\mrm{helicity}^i$, color $g\mrm{color}^i$ and antiparticle $g\mrm{anti}^i$ multiplicities
\begin{equation}
	\dfrac{\d^2 N_i}{\d t\d E} = \underbrace{g\mrm{color}^i\times g\mrm{helicity}^i \times g\mrm{anti}^i}_{\displaystyle g_i} \sum_{l,m} \dfrac{\d^2 N_{i,lm}}{\d t\d E}\,. \label{eq:emission_rate}
\end{equation}
In the case of a spherically symmetric metric, the sum over $m$ reduces to a factor $2l+1$ as all angular momentum projections give the same contribution to the total emission rate. The fields we consider in \texttt{BlackHawk} are the SM fields (photon, gluons, Higgs boson, W$^\pm$ boson, Z$^0$ boson, 3 neutrinos, electron, muon, tau, up, down, charm, strange, top, bottom quarks) plus possibly the graviton and some additional degree of freedom like a DM particle or dark radiation (DR). The multiplicities $g_i$, as well as the particles lifetimes and rest masses are summarized in Tables~\ref{tab:particles}, \ref{tab:BBN_particles} and \ref{tab:today_particles} in Appendix~\ref{app:particle_info}. To obtain the total primary spectrum of particle $i$ for an extended mass and/or secondary parameter $x$ distribution, we perform the integrals
\begin{equation}
	\dfrac{\d^2 n_i}{\d t\d E} = \int_{M\mrm{min}}^{M\mrm{max}} \d M \int_{x\mrm{min}}^{x\mrm{max}} \d x\,\dfrac{\d^2 N_i}{\d t\d E}\,\dfrac{\d n}{\d M} \, \dfrac{\d \tilde{n}}{\d x}\,. \label{eq:full_primary}
\end{equation}

\subsubsection{Black hole evolution}

Once we know the emission rates of all the particles of the (beyond the) SM spectrum, we can determine the evolution of the BH driven by Hawking evaporation. To do that, we need to obtain the differential equations of evolution of the BH mass and the possible secondary parameter. The mass loss rate is~\cite{Page:1976df,Page:1976ki,Page:1977um}
\begin{equation}
	f(M,x_j) \equiv  - M^2 \dfrac{\d M}{\d t} = M^2\int_{0}^{+\infty} E \sum_i \dfrac{\d^2 N_i}{\d t\d E} \,\d E\,, \label{eq:fMa}
\end{equation}
and the BH angular momentum loss rate is
\begin{equation}
	g(M,a^*) \equiv -\dfrac{M}{a^*} \dfrac{\d J}{\d t} = -\dfrac{M}{a^*}\int_{0}^{+\infty} \sum_i g_i \sum_{l,m} \dfrac{\d^2 N_{i,lm}}{\d t\d E} \,\d E\,. \label{eq:gMa}
\end{equation}
We could, on the same model, define the BH charge loss rate, but as we have not computed the emission of charged particles we can not get the full sum on the SM fields $i$. For higher-dimensional BHs, we have computed the emission rates on the 4-dimensional \textit{brane}, but not in the \textit{bulk} of the extra dimensions, where particles could acquire Kaluza-Klein piles of massive states. Thus, the sum is also incomplete in this case and we limit ourselves to the instantaneous emission. Finally, for polymerized BHs, only the factor $f(M,\varepsilon,a_0)$ is relevant as the secondary parameters $\varepsilon,a_0$ are constant over the BH evolution. As a convention, we can take $g = 0$ in this case. Then, we use Eqs.~\eqref{eq:fMa} and \eqref{eq:gMa} to get the differential equations for the BH mass and spin~\cite{Page:1976df,Page:1976ki,Page:1977um}
\begin{subequations}
	\begin{align}
		&\dfrac{\d M}{\d t} = - \dfrac{f(M,x_j)}{M^2}\,, \label{eq:dMdt} \\
		&\dfrac{\d a^*}{\d t} = a^*\dfrac{2f(M,a^*) - g(M,a^*)}{M^3}\,. \label{eq:dadt}
	\end{align}
\end{subequations}
The computation of the $f(M,x_j)$ and $g(M,a^*)$ Page factors in \texttt{BlackHawk} is described in Appendix~\ref{app:fM_tables}.

\subsubsection{Hadronization and decay}

The elementary particles emitted by BHs are not the final products of Hawking emission. Some of them are unstable, others only exist in composite states (hadrons). A particle physics code has to be used in order to evolve the elementary particles into final products. We used \texttt{PYTHIA}~\cite{Sjostrand:2014zea}, \texttt{HERWIG}~\cite{Bellm:2015jjp} and \texttt{Hazma}~\cite{Coogan:2019qpu} for this purpose.

The final particles, hereafter denoted as ``secondary Hawking particles'' (the elementary being the ``primary Hawking particles''), depend on the cosmological context in which they are emitted. For BBN studies, an estimation of the reaction rates imposes to consider only the particles with a lifetime longer than $\sim\!\!10^{-8}\,$s, listed in Table~\ref{tab:BBN_particles} of Appendix~\ref{app:particle_info}. For low redshift cosmological studies, one considers only the particles with an infinite lifetime, listed in Table~\ref{tab:today_particles}. The emission rate of secondary particle $j$ emitted by a distribution of BHs per units of time and energy is computed with the integral
\begin{equation}
	\dfrac{\d^2 N_j}{\d t\d E} = \int_0^{+\infty} \sum_{i} \dfrac{\d^2 N_i}{{\rm d}t{\rm d}E^\prime} \, \dfrac{\d N^i_j}{\d E}\, {\rm d}E^\prime\,, \label{eq:secondary}
\end{equation}
where the sum is taken over Hawking primary particles $i$. To obtain the total secondary spectrum of particle $j$ for an extended distribution of mass and/or secondary parameter $x$ we perform the integrals
\begin{equation}
	\dfrac{\d^2 n_j}{\d t\d E} = \int_{M\mrm{min}}^{M\mrm{max}} \d M \int_{x\mrm{min}}^{x\mrm{max}} \d x\,\dfrac{\d^2 N_j}{\d t\d E}\,\dfrac{\d n}{\d M} \, \dfrac{\d \tilde{n}}{\d x}\,. \label{eq:full_secondary}
\end{equation}
Appendix~\ref{app:hadronization_tables} describes how hadronization tables of the branching ratios $\d N_j^i(E^\prime,E)$ have been computed to transform the primary spectra into secondary spectra inside \texttt{BlackHawk}.


\section{Content and compilation}
\label{sec:content_compilation}

This Section describes the structure and file content of the code and explains its usage. \texttt{BlackHawk} is written in C and has been tested under Linux, Mac and Windows (using \texttt{Cygwin64}).

\subsection{Content}

\paragraph{\textbf{Main directory}} The main directory contains:
\begin{itemize}
	\item the source codes \texttt{BlackHawk\_inst.c} and \texttt{BlackHawk\_tot.c} containing the \texttt{main} routines,
	\item a pre-built parameter file \texttt{parameters.txt},
	\item a compilation file \texttt{Makefile},
	\item a \texttt{README.txt} file containing general information about the code,
	\item four folders \texttt{src/}, \texttt{results/}, \texttt{manual/} and \texttt{scripts/} that are described in the following.
\end{itemize}

\paragraph{\textbf{src/ subfolder}} This folder contains:
\begin{itemize}
	\item a header file \texttt{include.h} containing the declaration of all routines along with the parameter structure \texttt{struct param} (see Section~\ref{subsec:parameters_structure}) and the numerical values of general quantities (unit conversion factors, constants, particle masses, \textit{etc}),
	\item source files containing the definition of all the \texttt{BlackHawk} routines (\texttt{evolution.c}, \texttt{general.c}, \texttt{hadro\_herwig.c}, \texttt{hadro\_pythia.c}, \texttt{hadro\_pythianew.c}, \texttt{hadro\_hazma.c}, \texttt{primary.c}, \texttt{secondary.c}, \texttt{spectrum.c}, \texttt{technical.c}),
	\item compilation files \texttt{Makefile} and \texttt{FlagsForMake},
	\item a subfolder \texttt{tables/} containing all the numerical tables which will be described in the following.
\end{itemize}

\paragraph{\textbf{results/ subfolder}} This folder is designed to receive subfolders of data generated by running the \texttt{BlackHawk} code (see Section~\ref{sec:output_files}).

\paragraph{\textbf{manual/ subfolder}} This folder contains an up-to-date version of the present manual.

\paragraph{\textbf{scripts/ subfolder}} This folder contains all the scripts used to compute the numerical tables mentioned in the following, as well as visualization scripts and a main program for \texttt{SuperIso Relic}~\cite{Arbey:2009gu,Arbey:2011zz,Arbey:2018msw}. These scripts can be used to generate modified numerical tables for example (see Appendices~\ref{app:greybody_factors}, \ref{app:fM_tables} and \ref{app:hadronization_tables}). They are accompanied by \texttt{README.txt} files explaining how to use them.

\subsection{Compilation}

The compilation of \texttt{BlackHawk} has been tested on Linux, Mac and Windows (using \texttt{Cygwin64}) distributions. The code is written in \texttt{C99} standard. To compile the code, simply \texttt{cd} into the main directory and type\footnote{In case of problems of memory size at compilation, editing \texttt{src/include.h} and commenting \texttt{\#define HARDTABLES} can solve the problem at the price of a longer execution time.}:

\citecode{make BlackHawk\_*}

\noindent where \texttt{*} denotes \texttt{tot} or \texttt{inst}. This will create a library file \texttt{libblackhawk.a} and an executable file \texttt{BlackHawk\_*.x}. The compiler and compilation flags can be modified in \texttt{Makefile} if needed. To run the code, \texttt{cd} to the main directory and type\footnote{In case of memory problem at execution, increasing the stack size with the command \texttt{ulimit -s unlimited} can help solving the problem.}:

\citecode{./BlackHawk\_*.x parameter\_file}

\noindent where \texttt{parameter\_file} is the name of a parameter file (\textit{e.g.}~\texttt{parameters.txt} for the pre-built one). To compile only the library, just \texttt{cd} into the main directory and type:

\citecode{make}


\section{Input parameters}
\label{sec:input_parameters}

In this Section we describe how input parameters are handled in \texttt{BlackHawk} and their meaning.

\subsection{Parameter structure}
\label{subsec:parameters_structure}

The input parameters used by \texttt{BlackHawk} are listed in a parameter file (\textit{e.g.}~\texttt{parameters.txt} for the pre-built one). This file can be modified by the user and is saved for each new run of the code in the \texttt{results/} directory. A C structure has been defined in \texttt{include.h} to embed all the parameters (\textit{i.e.}~input parameters and default run parameters):\newline \newline
\texttt{
	struct param $\lbrace$\newline
	\indent char destination\_folder[200];\newline
	\indent int full\_output;\newline
	\indent int interpolation\_method;\newline \newline
	\indent int BH\_number;\newline
	\indent double Mmin;\newline
	\indent double Mmax;\newline
	\indent int metric;\newline
	\indent int param\_number;\newline
	\indent double amin;\newline
	\indent double amax;\newline
	\indent double Qmin;\newline
	\indent double Qmax;\newline
	\indent double epsilon\_LQG;\newline
	\indent double a0\_LQG;\newline
	\indent double n;\newline
	\indent double M\_star;\newline
	\indent int spectrum\_choice;\newline
	\indent int spectrum\_choice\_param;\newline
	\indent double amplitude;\newline
	\indent double stand\_dev;\newline
	\indent double crit\_mass;\newline
	\indent double eqstate;\newline
	\indent double stand\_dev\_param;\newline
	\indent double mean\_param;\newline
	\indent char table[32];\newline \newline
	\indent int tmin\_manual;\newline
	\indent double tmin;\newline
	\indent int limit;\newline
	\indent int nb\_fin\_times;\newline
	\indent int BH\_remnant;\newline
	\indent double M\_remnant;\newline \newline
	\indent int E\_number;\newline
	\indent double Emin;\newline
	\indent double Emax;\newline
	\indent int particle\_number;\newline
	\indent int grav;\newline
	\indent int add\_DM;\newline
	\indent double m\_DM;\newline
	\indent double spin\_DM;\newline
	\indent double dof\_DM;\newline \newline
	\indent int primary\_only;\newline
	\indent int hadronization\_choice;\newline \newline
	\indent double Mmin\_fM;\newline
	\indent double Mmax\_fM;\newline
	\indent int nb\_fM\_masses;\newline
	\indent int nb\_fM\_param;\newline
	\indent double param\_min;\newline
	\indent double param\_max;\newline
	\indent int nb\_gamma\_param;\newline
	\indent int nb\_gamma\_x;\newline
	\indent int nb\_gamma\_spins;\newline
	\indent int nb\_gamma\_fits;\newline
	\indent double Emin\_hadro;\newline
	\indent double Emax\_hadro;\newline
	\indent int nb\_init\_en;\newline
	\indent int nb\_fin\_en;\newline
	\indent int nb\_init\_part;\newline
	\indent int nb\_fin\_part;\newline
	$\rbrace$;\newline
}

Most routines described in Section~\ref{sec:routines} will use this structure as an argument in order to have an easy access to the run parameters. Depending on the choices of the parameters, some of them can be irrelevant for a given run and will therefore not be taken into account, and no error message will be displayed for the irrelevant/unused parameters. Error messages are displayed if some parameter is out of its allowed range, automatically cancelling the run.

\subsection{General parameters}

This set of parameters defines the general variables:
\begin{itemize}
	\item \texttt{destination\_folder} is the name of the output folder that will be created in \texttt{results/} to save run data.
	\item \texttt{full\_output} determines whether the shell output will be expanded (\texttt{full\-\_output = 1}) or not (\texttt{full\_output = 0}). It can be useful to debug the code or to follow the progress in time-consuming routines. It also determines whether some interactive pre-run checks are done with the user. We thus recommend that this parameter is always set to 1.
	\item \texttt{interpolation\_method} determines whether the interpolations in the numerical tables are made linearly (interpolation between the tabulated values) or logarithmically (linear interpolation between the decimal logarithm of the tabulated values).
\end{itemize}

\subsection{BH spectrum parameters}
\label{params:spectrum}

This set of parameters defines the quantities used to compute the BH density distribution:
\begin{itemize}
	\item \texttt{BH\_number} is the number of BH masses that will be simulated. If the parameter \texttt{spectrum\_choice} is not set to $-1$, it has to be an integer greater than or equal to 1. If it is equal to 1, the only BH mass will be \texttt{Mmin} (see below). If the parameter \texttt{spectrum\_choice} is set to $-1$, it has to be the number of tabulated values in the user-defined BH distribution. It will be automatically set to 1 if \texttt{spectrum\_choice} is set to 0 (Dirac distribution).
	
	\item \texttt{Mmin} and \texttt{Mmax} are respectively the lowest and highest BH masses that will be simulated. They have to be given in grams and satisfy the condition $M\mrm{p} \approx 2\times10^{-5}\,$g $<$ \texttt{Mmin}, \texttt{Mmax}, where $M\mrm{p}$ is the Planck mass. For an extended mass distribution, one must have \texttt{Mmin} $<$ \texttt{Mmax}. If they are not compatible with boundaries of the mass distribution, the computation will stop.
	
	\item \texttt{metric} switches between the different BH solutions available in \texttt{BlackHawk}: 0 (Kerr BH), 1 (polymerized BH), 2 (Reissner--Nordstr\"om BH), 3 (higher-dimensional BH).
	
	\item \texttt{param\_number} is the number of BH secondary parameters that will be simulated. It is relevant only if \texttt{metric} is 0 or 2, otherwise there is only one possible value of the secondary parameters. If the parameter \texttt{spectrum\_choice} is not set to $-1$, it has to be an integer greater than or equal to 1. If it is equal to 1, the only BH secondary parameter will be \texttt{amin} (Kerr BHs), \texttt{Qmin} (Reissner--Nordstr\"om BHs) or the single-valued secondary parameter in the case of polymerized or higher-dimensional BHs. If the parameter \texttt{spectrum\_choice} is set to $-1$, it has to be the number of tabulated values in the user-defined BH distribution. It will be automatically set to 1 if \texttt{spectrum\_choice\_param} is set to 0 or if \texttt{metric} is set to 1 or 2.
	
	\item \texttt{amin} and \texttt{amax} are respectively the lowest and highest BH dimensionless reduced spins that will be simulated. They have to satisfy the condition $0\le$ \texttt{amin}, \texttt{amax} $<1$. For an extended spin distribution, one must have \texttt{amin} $<$ \texttt{amax}. If they are not compatible with boundaries of the spin distribution, the computation will stop.	The paradigm is the same with \texttt{Qmin} and \texttt{Qmax} for the dimensionless BH electric charge.
	
	\item \texttt{epsilon\_LQG} is the dimensionless quantization parameter $\varepsilon$ for polymerized BHs. It has to be a positive number $0\le\varepsilon\le 100$; \texttt{a0\_LQG} is the area gap $a_0$ in loop quantum gravity, it has to be a positive number, but only the two values 0 and $0.11$ have been used to compute the numerical tables.
	
	\item \texttt{n} is the number of extra spatial dimensions $n$ for higher-dimensional BHs, it has to be an integer $0 \le n \le 6$; \texttt{M\_star} is the rescaled Planck mass in Eq.~\eqref{eq:rH_higher}, only the value $M_* = 1$ has been used in the numerical tables.
	
	\item \texttt{spectrum\_choice} selects the form of the BH mass distribution, it has to be an integer among 0 (Dirac), 1 (log-normal for the mass density), 11 (log-normal for the number density), 2 (power-law), 3 (critical collapse), 4 (peak theory), 5 (uniform) and $-1$ (user defined distribution, provided as a \texttt{.txt} file in the subfolder \texttt{/src/tables/users\_spectra/}).
	
	\item \texttt{spectrum\_choice\_param} selects the form of the BH secondary parameter distribution. It has to be an integer among 0 (Dirac), 1 (uniform) and 2 (Gaussian). If the parameter \texttt{spectrum\_choice} has been set to $-1$, then the spin distribution is also the user-defined one and this parameter is irrelevant. If the BH metric is set to polymerized or higher-dimensional BHs, then the secondary parameter is automatically single-valued.
	
	\item \texttt{amplitude} is the amplitude $A$ present in Eqs.~(\ref{eq:lognormal}), (\ref{eq:lognormal2}), (\ref{eq:powerlaw}), (\ref{eq:criticalcollapse}) and (\ref{eq:uniform_M}). It is the normalization of the corresponding BH distribution and thus strictly positive, its unit depends on the distribution chosen, but masses are in grams and densities in centimetres$^{-3}$.
	
	\item \texttt{stand\_dev} is the dimensionless standard deviation $\sigma$ in the log-normal distributions of Eqs.~(\ref{eq:lognormal}) and (\ref{eq:lognormal2}). It has to be strictly positive.
	
	\item \texttt{crit\_mass} is the characteristic mass $M\mrm{c}$ in Eqs.~(\ref{eq:lognormal}) and (\ref{eq:lognormal2}) and $M\mrm{f}$ in Eq.~\eqref{eq:criticalcollapse}. It has to be strictly positive and given in grams.
	
	\item \texttt{eq\_state} defines the equation of state $w$ for the power-law mass distribution of Eq.~\eqref{eq:powerlaw}, it is dimensionless.
	
	\item \texttt{stand\_dev\_param} is the dimensionless standard deviation $\sigma$ in the Gaussian distribution~\eqref{eq:gaussian_param} of the secondary parameter; it has to be strictly positive.
	
	\item \texttt{mean\_param} is the characteristic dimensionless secondary parameter $x\mrm{c}$ in the Gaussian distribution~\eqref{eq:gaussian_param} or in the uniform distribution~\eqref{eq:uniform_param}, it has to be strictly positive.
	
	\item \texttt{table} is the name of a user-defined BH distribution table. It has to be a string with any file extension. This file has to have the same format as the \texttt{BlackHawk} generated files, that is to say a two-entry table with lines corresponding to the BH mass and rows to the BH secondary parameters.
\end{itemize}
If the parameters \texttt{spectrum\_choice} and \texttt{spectrum\_choice\_param} are both set to 0, then the distribution is monochromatic in both mass and secondary parameter, mimicking a single BH --- that is to say, the emissivities obtained are those of a single BH. This option can be useful to compute and compare known test emissivities of single BHs.

\subsection{BH evolution parameters}

This set of parameters defines the quantities used to compute the BH evolution:
\begin{itemize}
	\item \texttt{tmin\_manual} switches between automatically set $t\mrm{min}$ for the formation time of BHs (\texttt{tmin\_manual = 0}) following Eq.~\eqref{eq:formation_time} and manually set $t\mrm{min}$ (\texttt{tmin\_manual = 1}). For now, if the BH distribution is extended in mass, then all BHs are formed at the same time --- meaning, the formation time is common to all the initial BH masses. This should have negligible impact on the emission rates.
	
	\item \texttt{tmin} is the initial integration time of the evolution of BH, in seconds, if \texttt{tmin\_manual} has been set to 1. It can have any positive value.
	
	\item \texttt{limit} is the iteration limit when computing the time evolution of a single BH. It is fixed to \texttt{limit = 5000} even if the effective iteration number hardly reaches 1000. It should be increased if the integration does not reach the complete evaporation of BHs or if the following error appears:
	
	\citecode{[life\_evolution] : ERROR ITERATION LIMIT REACHED !}
	
	\item \texttt{nb\_fin\_times} is the number of final integration times of the computations. It is set automatically by the integration procedure.
	
	\item \texttt{BH\_remnant} switches between two cases: either the BH evaporation continues until the Planck mass is reached, and at this moment we are agnostic about what happens but we believe that Hawking radiation becomes irrelevant (\texttt{BH\_remnant = 0}); or the BH evaporation stops before reaching the Planck mass, leaving a massive remnant (\texttt{BH\_remnant = 1}).
	
	\item \texttt{M\_remnant} is the BH remnant mass in grams if \texttt{BH\_remnant = 1}.
\end{itemize}

\subsection{Primary spectrum parameters}

This set of parameters defines the quantities related to the primary Hawking spectra:
\begin{itemize}
	\item \texttt{E\_number} is the number of primary particle energies that will be simulated. It has to be an integer greater than or equal to 2.
	
	\item \texttt{Emin} and \texttt{Emax} are the minimum and maximum primary particle energies, respectively. They must be compatible with the numerical tables boundaries and satisfy $0 < \texttt{Emin} < \texttt{Emax}$.
	
	\item \texttt{particle\_number} is the number of primary particle types. It is fixed to 15 (photon, gluon, W$^\pm$ boson, Z$^0$ boson, Higgs boson, neutrino, 3 leptons (electron, muon, tau) and 6 quarks (up, down, charm, strange, top, bottom)), but is increased in the code if graviton ($+1$) or DM emission ($+1$) is activated.
	
	\item \texttt{grav} determines whether the emission of gravitons by BHs will be taken into account (\texttt{grav = 1}) or not (\texttt{grav = 0}).
	
	\item \texttt{add\_DM} swithces between DM emission (\texttt{add\_DM = 1}) or not (\texttt{add\_DM = 0}). It is automatically set to 1 if \texttt{hadronization\_choice} is set to 3 (\texttt{Hazma} tables), as in this case the DM slot will be used to compute the emission of primary pions, thus forbidding a mixed scenario ``DM + \texttt{Hazma} hadronization''.
	
	\item \texttt{m\_DM}, \texttt{spin\_DM} and \texttt{dof\_DM} are respectively the additional DM mass in GeV, DM spin (among 0, 1, 2 and $1/2$, while the value $3/2$ is only available for Schwarzschild BHs as these are the only spin $3/2$ greybody factors we have computed) and DM number of dof (any positive integer). The evolution numerical tables $f(M,x_j)$ and $g(M,a^*)$ have been computed for \texttt{dof\_DM = 1} only, but the change should be negligible if \texttt{dof\_DM} remains much smaller than the number of dof available at some BH mass $M$ ($g = 106.75$ at $M\ll 10^{10}\,$g for the full SM, $g = 7.25$ at $M\gg 10^{19}\,$g for the photon and 3 massless Majorana neutrinos); the user can recompute these tables anyway if needed.
\end{itemize}

\subsection{Hadronization parameters}

This set of parameters defines the quantities used during hadronization:
\begin{itemize}
	\item \texttt{primary\_only} determines whether the secondary spectra will be computed or not. It has to be an integer between 0 (primary and secondary spectra) and 1 (primary spectra only). In the case where the parameters \texttt{Emin} and \texttt{Emax} are not compatible with the hadronization tables boundaries, a warning will be displayed and extrapolation used (\texttt{hadronization\_choice = 0,1,2}), or the computation will stop altogether (\texttt{hadronization\_choice = 3}).
	
	\item \texttt{hadronization\_choice} determines which hadronization tables will be used to compute the secondary spectra. It has to be an integer between 0 (\texttt{PYTHIA} tables, early Universe/BBN epoch), 1 (\texttt{HERWIG} tables, early Universe/BBN epoch), 2 (\texttt{PYTHIA} tables, present epoch) and 3 (\texttt{Hazma} tables, present epoch).
\end{itemize}

\subsection{Tables parameters}

This set of parameters is quite special: it regroups the numerical table parameters, which are fixed when the tables are computed and should not be modified by the user. Thus, they are set in separated \texttt{infos.txt} files. Their list is:
\begin{itemize}
	\item \texttt{Mmin\_fM}, \texttt{Mmax\_fM}, \texttt{nb\_fM\_masses} and \texttt{nb\_fM\_param} describe the format of the evolution tables $f(M,x_j)$ and $g(M,a^*)$ (see Appendix~\ref{app:fM_tables}). They are read in the file:
	
	\citecode{src/tables/fM\_tables/infos.txt}
	
	\item \texttt{param\_min}, \texttt{param\_max}, \texttt{nb\_gamma\_param}, \texttt{nb\_gamma\_x}, \texttt{nb\_gamma\_spins} and \texttt{nb\_gamma\_fits} describe the format of the greybody factor tables (see Appendix~\ref{app:greybody_factors}). They are read in the file:
	
	\citecode{src/tables/gamma\_tables/infos.txt}
	
	\item \texttt{Emin\_hadro}, \texttt{Emax\_hadro}, \texttt{nb\_init\_en}, \texttt{nb\_fin\_en}, \texttt{nb\_init\_part} and \texttt{nb\_fin\_part} describe the format of the hadronization tables (see Appendix~\ref{app:hadronization_tables}). They are read in the file:
	
	\citecode{src/tables/hadronization\_tables/infos.txt}
\end{itemize}


\section{Routines}
\label{sec:routines}

In this Section, we have listed all the routines defined in \texttt{BlackHawk}. To simplify the analytic formulas, all intermediate quantities are in GeV (see Appendix~\ref{app:units} for conversion rules).

\subsection{General routines}
\label{rout:general}

There are seven general routines in the \texttt{BlackHawk} code. The principal ones are the two \texttt{main} routines, described in Section~\ref{sec:programs}. The other five are:
\begin{itemize}
	\item \texttt{int read\_params(struct param *parameters, char name[], int session)}\newline
	This routine reads the file \texttt{name} and the numerical tables information files thanks to the \texttt{read\_*\_infos} routines. The parameters are converted from CGS units to GeV. The user should respect the original syntax when modifying the parameters (concerning spaces, underscores, ...), except for comments which are preceded by a \texttt{\#} symbol and are skipped automatically. It takes a pointer to a \texttt{struct param} object (see Section~\ref{subsec:parameters_structure}) as an argument and fills it using the file \texttt{name}. The argument \texttt{session} keeps track of which one of the main programs has been launched (0 for \texttt{BlackHawk\_tot}, 1 for \texttt{BlackHawk\_inst}). If one parameter is not of the type described in Section~\ref{sec:input_parameters} this function will display an error message of the kind:
	
	\citecode{[read\_params] : ERROR ... !}
	
	\noindent sometimes accompanied with an explanation message. Any of these errors will make the routine return the value 0 and end the \texttt{BlackHawk} run. If one parameter is in contradiction with the others but the computation can still be partly done (\textit{e.g.}~only the primary spectra can be computed with the given parameters), a warning message will be displayed, of the kind:
	
	\citecode{[read\_params] : WARNING ... !}
	
	\noindent In such a case, the problematic parameters will be set automatically (\textit{e.g.}~\texttt{primary\_only = 1}) and the computation will be performed.
	
	\item \texttt{int read\_fM\_infos(struct param *parameters)}\newline
	This routine reads the $f(M,x_j)$ and $g(M,a^*)$ table information in the file:
	
	\citecode{src/tables/fM\_tables/infos.txt}
	
	\noindent It returns 1 if all parameters are correct and 0 if there is an error, displaying an error message in this case:
	
	\citecode{[read\_fM\_infos] : ERROR ... !}
	
	\item \texttt{int read\_gamma\_infos(struct param *parameters)}\newline
	This routine reads the greybody factor table information in the file:
	
	\citecode{src/tables/gamma\_tables/infos.txt}
	
	\noindent It returns 1 if all parameters are correct and 0 if there is an error, displaying a message in this case:
	
	\citecode{[read\_gamma\_infos] : ERROR ... !}
	
	\item \texttt{int read\_hadronization\_infos(struct param *parameters)}\newline
	This routine reads the hadronization table information in the file:
	
	\citecode{src/tables/hadronization\_tables/infos.txt}
	
	\noindent It returns 1 if all parameters are correct and 0 if there is an error, displaying a message in this case:
	
	\citecode{[read\_hadronization\_infos] : ERROR ... !}
	
	\item \texttt{int memory\_estimation(struct param *parameters, int session)}\newline
	This routine gives a rough estimate of the usage of both RAM and disk space (see Section~\ref{sec:memory_usage}). If the parameter \texttt{full\_output} is set to 1, the user has the choice to cancel the run and the value 0 is returned, otherwise it is 1. The output is given in MB.
\end{itemize}

\subsection{BH initial distribution routines}
\label{rout:spectrum}

There are six routines contributing to the BH initial distribution computation:
\begin{itemize}
	\item \texttt{void read\_users\_table(double *init\_masses, double *init\_params, double **spec\_table, struct\\ param *parameters)}\newline
	This routine reads a user-defined BH distribution table in the file given by the parameter \texttt{table}, if the parameter \texttt{spectrum\_choice} is set to $-1$. It fills the arrays \texttt{init\_masses[]}, \texttt{init\_params[]} and \texttt{spec\_table[][]} with results converted from CGS units to GeV.
	
	\item \texttt{double nu(double M)}\newline
	This routine takes a BH mass as an argument and computes the dimensionless quantity $\nu(M)$ defined in Eq.~\eqref{eq:nu}.
	
	\item \texttt{double M\_dist(double M, struct param *parameters)}\newline
	This routine takes a BH mass as an argument and computes the comoving density $\d n/\d M$ defined in Eq.~(\ref{eq:peak_theory}) (using the \texttt{nu} routine) or Eqs.~(\ref{eq:lognormal}), \eqref{eq:lognormal2}, (\ref{eq:powerlaw}), (\ref{eq:criticalcollapse}) or (\ref{eq:uniform_M}) (in GeV$^{2}$ $\rightarrow$ cm$^{-3}\cdot$g$^{-1}$), depending on the parameter \texttt{spectrum\_choice}. If this parameter is set to 0, a flat distribution is used with only one BH mass.
	
	\item \texttt{double param\_dist(double param, struct param *parameters)}\newline
	This routine takes a BH secondary parameter as an argument ($x = a^*$ or $x = Q^*$) and computes the fraction $\d\tilde{n}/\d x$ defined in Eqs.~(\ref{eq:gaussian_param}) or (\ref{eq:uniform_param}) (dimensionless), depending on the parameter \texttt{spectrum\_choice\_param}. If this parameter is set to 0, a Dirac distribution is used with only one $x$ value.
	
	\item \texttt{void spectrum(double *init\_masses, double *init\_spins, double **spec\_table, struct param\\ *parameters)}\newline
	This routine fills the array \texttt{init\_masses[]} with \texttt{BH\_number} BH masses logarithmically distributed between \texttt{Mmin} and \texttt{Mmax}. If the parameter \texttt{BH\_number} is set to 1, the only BH initial mass will be \texttt{Mmin}. It then fills the array \texttt{init\_params[]} with secondary parameters linearly distributed between $a\mrm{min},a\mrm{max}$ or $Q\mrm{min},Q\mrm{max}$ (for Kerr and Reissner--Norstr\"om BHs respectively), and finally fills the array \texttt{spec\_tables[][]} computing the corresponding comoving densities ${\rm d}n(M,x)$ (in GeV$^3$ $\rightarrow$ cm$^{-3}$) using the \texttt{M\_dist} and \texttt{param\_dist} routines where ${\rm d}M$ is taken around the considered mass and ${\rm d}x$ around the considered secondary parameter. The result is rescaled by a factor $10^{100}$ due to the very small numbers involved.
	
	\item \texttt{void write\_spectrum(double *init\_masses, double *init\_spins, double **spec\_table, struct param *parameters)}\newline
	This routine writes the BH initial masses, secondary parameters and comoving densities in the file:
	
	\citecode{results/destination\_folder/BH\_spectrum.txt}
	
	detailed in Section~\ref{sec:output_files}. The results are converted from GeV to CGS units.
\end{itemize}

\subsection{BH evolution routines}
\label{rout:life}

There are nineteen routines contributing to the BH time evolution computation:
\begin{itemize}
	\item \texttt{double rplus\_Kerr(double M, double a)}\newline
	This routine gives the external Kerr radius $r_+ \equiv r\mrm{H}$ of Eq.~\eqref{eq:rH_Kerr} of a rotating BH for a given mass $M$ and reduced spin $a^*$ (in GeV$^{-1}$ $\rightarrow$ cm).
	
	\item \texttt{double temp\_Kerr(double M, double a)}\newline
	This routine gives the Hawking temperature of a Kerr BH for a given mass $M$ and reduced spin $a^*$, see Table~\ref{tab:temperatures} (in GeV $\rightarrow$ K).
	
	\item \texttt{double P\_LQG(double epsilon)}\newline
	This routine computes the dimensionless polymerization function $P(\varepsilon)$ defined in Eq.~\eqref{eq:poly_function}.
	
	\item \texttt{double m\_LQG(double M, double epsilon)}\newline
	This routine computes the effective mass $m(\varepsilon)$ of a polymerized BH with ADM mass $M$, defined in Eq.~\eqref{eq:m_LQG} (in GeV $\rightarrow$ g).
	
	\item \texttt{double temp\_LQG(double M, double epsilon, double a0)}\newline
	This routine gives the temperature of a polymerized BH of mass $M$, polymerization factor $\varepsilon$ and area gap $a_0$, see Table~\ref{tab:temperatures} (in GeV $\rightarrow$ K).
	
	\item \texttt{double rplus\_charged(double M, double Q)} and \texttt{rminus\_charged(double M, double Q)}\newline
	These routines compute the external horizon and Cauchy radii $r\mrm{H} \equiv r_+$ and $r\mrm{C} \equiv r_-$ of a Reissner--Nordstr\"om BH of mass $M$ and adimensioned charge $Q^*$, defined in Eq.~\eqref{eq:rH_charged} (in GeV$^{-1}$ $\rightarrow$ cm).
	
	\item \texttt{double temp\_charged(double M, double Q)}\newline
	This routine gives the temperature of a Reissner--Nordstr\"om BH of mass $M$ and adimensioned charge $Q^*$, see Table~\ref{tab:temperatures} (in GeV $\rightarrow$ K).
	
	\item \texttt{double rH\_higher(double M, double n, double M\_star)}\newline
	This routine computes the horizon radius of a higher-dimensional BH of mass $M$ and extra dimensions number $n$ with rescaled Planck mass $M_*$, defined in Eq.~\eqref{eq:rH_higher} (in GeV$^{-1}$ $\rightarrow$ cm).
	
	\item \texttt{double temp\_higher(double M, double n, double M\_star)}\newline
	This routine gives the temperature of a higher-dimensional BH of mass $M$ and extra dimensions number $n$ with rescaled Planck mass $M_*$, see Table~\ref{tab:temperatures} (in GeV $\rightarrow$ K).
	
	\item \texttt{void read\_fM\_table(double **fM\_table, double *fM\_masses, double *fM\_param, struct param\\ *parameters)}\newline
	This routine reads the $f(M,x_j)$ factors defined in Eq.~\eqref{eq:fMa} in the relevant table contained in the folder:
	
	\citecode{src/tables/fM\_tables/}
	
	\noindent depending on the parameter \texttt{metric}. It fills the arrays \texttt{fM\_masses[]} (in GeV $\rightarrow$ g), \texttt{fM\_param[]} (dimensionless) and \texttt{fM\_table[][]} with format [mass][param] (in GeV$^{4}$ $\rightarrow$ g$^3\cdot$s$^{-1}$).
	
	\item \texttt{void read\_gM\_table(double **gM\_table, double *fM\_masses, double	*fM\_param, struct param\\ *parameters)}\newline
	This routine reads the $g(M,a^*)$ factor defined in Eq.~\eqref{eq:gMa} in the relevant table contained in the folder:
	
	\citecode{src/tables/fM\_tables/}
	
	\noindent It fills the arrays \texttt{fM\_masses[]} (in GeV $\rightarrow$ g), \texttt{fM\_param[]} and \texttt{gM\_table[][]} with format [mass][param] (in GeV$^{4}$ $\rightarrow$ g$^2\cdot$GeV$\cdot$s$^{-1}$).
	
	\item \texttt{double loss\_rate\_M(double M, double param, double **fM\_table, double *fM\_masses, double\\ *fM\_param, int counter\_M, int counter\_param, struct param *parameters)}\newline
	This routine computes the quantity $\d M/\d t$ defined in Eq.~\eqref{eq:dMdt} (in GeV$^2$ $\rightarrow$ g$\cdot$s$^{-1}$).
	
	\item \texttt{double loss\_rate\_param(double M, double param, double **fM\_table, double **gM\_table, double\\ *fM\_masses, double *fM\_param, int counter\_M, int counter\_param, struct param *parameters)}\newline
	This routine computes the quantity $\d a^*/\d t$ defined in Eq.~\eqref{eq:dadt} (in GeV $\rightarrow$ s$^{-1}$).
	
	\item \texttt{evolution\_times(double *init\_masses, double *init\_params, double **evol\_times, double\\ **fM\_table, double **gM\_table, double *fM\_masses, double *fM\_a,	struct param *parameters)}\newline
	This routine computes the lifetime of all BHs with masses in \texttt{init\_masses[]} and secondary parameters in \texttt{init\_params[]} and stores the results in \texttt{evol\_times[][]} with format [mass][param] (in GeV$^{-1}$ $\rightarrow$ s).
	
	\item \texttt{sort\_lifetimes(double **evol\_times, double *sorted\_times, int **rank, struct param\\ *parameters)}\newline
	This routine sorts the lifetimes of the BHs computed by \texttt{evolution\_times} from shortest to longest using the routine \texttt{sort\_fusion} and stores the corresponding ranks in the table \texttt{rank[][]} with format [mass][param].
	
	\item \texttt{void life\_evolution(double ***life\_masses, double ***life\_params, double *life\_times, double *dts, double *init\_masses, double *init\_params, int **rank, double **fM\_table, double\\ **gM\_table, double *fM\_masses, double *fM\_param, struct param *parameters)}\newline
	This routine computes the evolution of each of the initial BH masses in \texttt{init\_masses[]} and BH secondary parameters in \texttt{init\_params[]}. The initial time \texttt{life\_times[0]} is set to \texttt{tmin}, the initial masses \texttt{life\_masses[i][j][0]} are set to \texttt{init\_masses[i]} and the initial secondary parameters \texttt{life\_params[i][j][0]} are set to \texttt{init\_params[j]}. Iteratively, the next masses and secondary parameters are estimated using the simple Euler method
	\begin{subequations}
		\begin{align}
			&M(t+{\rm d}t) = M(t) + \dfrac{{\rm d}M}{{\rm d}t}{\rm d}t\,, \\
			&x(t + {\rm d}t) = x(t) + \dfrac{{\rm d}x}{{\rm d}t}{\rm d}t\,,
		\end{align}
	\end{subequations}
	where the derivatives are computed using the \texttt{loss\_rate\_*} routines. If the mass or the secondary parameter relative variations of the currently interesting BH are too large ($|{\rm d}X/X| > 0.1$) then the time interval is divided by 2. If all the variations are very small ($|{\rm d}X/X| < 0.001$), and if the current timestep is reasonable compared to the current timescale (${\rm d}t/t\lesssim 1$) then the time interval is multiplied by 2. If the dimensionless spin (Kerr BHs) reaches $10^{-3}$, we stop computing its variation and simply set it to 0, and it does not enter any more in the adaptive timesteps conditions. This goes on until each BH reaches the Planck mass or the remnant mass. As BH evolution is very steep towards the end of evaporation, the BH that determines the time interval is determined by following the order given by the array \texttt{rank[][]} which contains the information about the chronological order in which the BHs will evaporate. If the recursion limit \texttt{limit} $\times$ \texttt{BH\_number} is reached, the following error is displayed:
	
	\citecode{[life\_evolution] : ERROR ITERATION LIMIT REACHED !}
	
	\noindent This may be a sign that the parameter \texttt{limit} should be increased. The intermediate time intervals ${\rm d}t$, times $t$, masses $M$ and secondary parameters $x$ are stored in the arrays \texttt{dts[]}, \texttt{life\_times[]} (both in GeV$^{-1}$ $\rightarrow$ s), \texttt{life\_masses[][][]} (in GeV $\rightarrow$ g) and \texttt{life\_params[][][]} (dimensionless), respectively, with format [mass][param][time] for the two last ones.
	
	\item \texttt{void write\_life\_evolutions(double ***life\_masses, double ***life\_params, double *life\_times, struct param *parameters)}\newline
	This routine writes the BH time-dependent masses and secondary parameters until the end of evaporation in the file:
	
	\citecode{results/destination\_folder/life\_evolutions.txt}
	
	detailed in Section~\ref{sec:output_files}. If the BH distribution is extended in both mass and secondary parameter, we advise the user to deactivate this writing because of the extensive memory space used in this case (see Section~\ref{sec:memory_usage}). The results are converted from GeV to CGS units.
\end{itemize}

\subsection{Primary spectra routines}
\label{rout:primary}

There are five routines contributing to the computation of the primary Hawking spectra:
\begin{itemize}
	\item \texttt{void read\_gamma\_tables(double ***gammas, double *gamma\_param, double *gamma\_x, struct param\\ *parameters)}\newline
	This routine reads the greybody factors $\Gamma_{s_i l m}(E,M,x_j)/(e^{E^\prime/T}  - (-1)^{2s_i})$ defined in Eq.~\eqref{eq:hawking_master}, in the files:
	
	\citecode{src/tables/gamma\_tables/[*/]spin\_*.txt}
	
	\noindent It fills the arrays \texttt{gamma\_param[]} and \texttt{gamma\_x[]} with the tabulated secondary parameters (dimensionless) and $x\equiv Er\mrm{S}$ (dimensionless $\rightarrow$ GeV$\cdot$cm), respectively. It fills the array \texttt{gammas[][][]} with the corresponding dimensionless greybody factors in format [type][param][$x$].
	
	\item \texttt{void read\_asymp\_fits(double ***fits, struct param *parameters)}\newline
	This routine reads the asymptotic fit parameters for the greybody factors, contained in the files:
	
	\citecode{src/tables/gamma\_tables/[*/]spin\_*\_fits.txt}
	
	\noindent It fills the array \texttt{fits[][][]} with format [type][param][coefs].
	
	\item \texttt{double dNdtdE(double E, double M, double param, int particle\_index, double ***gammas, double *gamma\_param, double *gamma\_x, double ***fits, double *dof, double *spins, double\\ *masses\_primary, int counter\_param, int counter\_x, struct param *parameters)}\newline
	This routine computes the emission rate ${\rm d}^2N_i/{\rm d}t{\rm d}E$, defined in Eq.~\eqref{eq:emission_rate} of the primary particle \texttt{particle\_index}, for a given particle energy $E$, BH mass $M$, secondary parameter and with the primary particle information contained in the arrays \texttt{dof[]}, \texttt{spins[]} and \texttt{masses\_primary[]}. If $x \equiv Er\mrm{S}$ is inside the greybody factor tables boundaries, the values are interpolated in these tables at position \texttt{counter\_param} and \texttt{counter\_x}. Otherwise, it uses the asymptotic high and low energy fit tables. If the energy condition $E>\mu$ is broken, where $\mu$ is the particle rest mass, it returns 0. The result is dimensionless ($\rightarrow$ GeV$^{-1}\cdot$s$^{-1}$).
	
	\item \texttt{void instantaneous\_primary\_spectrum(double **instantaneous\_primary\_spectra, double *BH\_masses, double *BH\_params, double **spec\_table, double *energies, double ***gammas, double\\ *gamma\_param, double *gamma\_x, double ***fits, double *dof, double *spins, double\\ *masses\_primary, struct param *parameters)}\newline
	This routine computes the instantaneous primary Hawking spectra for a distribution of BHs given by the routine \texttt{spectrum}, namely the quantities $\d^2 n_i/\d t\d E$ defined in Eq.~\eqref{eq:full_primary} for each primary particle $i$ and each primary energy in \texttt{energies[]}, thanks to the routine \texttt{dNdtdE}. The results are stored in the array \texttt{instantaneous\_primary\_spectra[][]} with format [particle][energy].
	
	\item \texttt{void write\_instantaneous\_primary\_spectra(double **instantaneous\_primary\_spectra, double\\ *energies, struct param *parameters)}\newline
	This routine writes the instantaneous primary Hawking spectra in the file:
	
	\citecode{results/destination\_folder/instantaneous\_primary\_spectra.txt}
	
	\noindent detailed in Section~\ref{sec:output_files}. The results are converted from GeV to CGS units.
\end{itemize}

\subsection{Secondary spectrum routines}
\label{rout:secondary}

There are thirteen routines contributing to the computation of the secondary Hawking spectra:
\begin{itemize}
	\item \texttt{void convert\_hadronization\_tables(double ****tables, double *initial\_energies, double\\ *final\_energies, struct param *parameters)}\newline
	This routine is auxiliary. It writes hardcoded versions of the hadronization tables in files:
	
	\citecode{src/tables/hadronization\_tables/hadronization\_tables\_*.h}
	
	\noindent in order to accelerate the code execution (while slightly slowing its compilation).
	
	\item \texttt{void read\_hadronization\_tables(double ****tables, double *initial\_energies, double\\ *final\_energies, struct param *parameters)}\newline
	This routine reads the relevant hadronization tables in the files:
	
	\citecode{src/tables/hadronization\_tables/*\_tables/*.txt}
	
	depending on the parameter \texttt{hadronization\_choice}. If \texttt{HARDTABLES} is defined, it uses the hardcoded tables included at compilation thanks to the routines \texttt{read\_hadronization\_*}, which is faster. It fills the arrays \texttt{initial\_energies[]} and \texttt{final\_energies[]} with the tabulated primary particles and secondary particles energies (in GeV), respectively, and fills the array \texttt{tables[][][][]} with the branching ratios $\d N_j^i/\d E$ defined in Eq.~\eqref{eq:secondary} (in GeV$^{-1}$) with format [secondary particle][initial energy][final energy][primary particle].
	
	\item \texttt{void total\_spectra(double ***partial\_hadronized\_spectra, double **partial\_primary\_spectra, \\double **partial\_integrated\_hadronized\_spectra, double ****tables, double *initial\_energies, double *final\_energies, double ***primary\_spectra, double *times, double *energies, double\\ *masses\_secondary, struct param *parameters)}\newline
	This routine is a container that uses the ``instantaneous'' routines to compute the Hawking primary and secondary spectra at each timestep in \texttt{times} and writes it directly in the output in order to save RAM memory. To do so, it creates the output files (one per particle)
	
	\citecode{results/destination\_folder/*\_primary\_spectrum.txt}
	
	\noindent and
	
	\citecode{results/destination\_folder/*\_secondary\_spectrum.txt}
	
	\noindent if \texttt{primary\_only} is set to 0. It reads the writing instructions using \texttt{read\_writing\_instructions}. Then, it fills the partial arrays \texttt{partial\_*} with the instantaneous primary spectra, hadronized spectra and integrated spectra at each intermediate time and calls the routine \texttt{write\_lines} to write the partial result in the output before moving to the next timestep.
	
	\item \texttt{void read\_writing\_instructions(int *write\_primary, int *write\_secondary, struct param\\ *parameters)}\newline
	This routine reads the writing information contained in the files
	
	\citecode{src/tables/write\_*.txt}
	
	depending on the \texttt{hadronization\_choice}, and stores it into the arrays \texttt{write\_primary[]} and \texttt{write\_secondary[]}.
	
	\item \texttt{void write\_lines(char **file\_names, double **partial\_integrated\_hadronized\_spectra, int\\ *write\_primary, int *write\_secondary, double time, struct param *parameters)}\newline
	This routine writes, given a time $t$ and instantaneous primary and secondary spectra (if \texttt{primary\_only} is set to 0), a new line in the files:
	
	\citecode{results/destination\_folder/*\_primary\_spectrum.txt}
	
	and
	
	\citecode{results/destination\_folder/*\_secondary\_spectrum.txt}
	
	\noindent detailed in Section~\ref{sec:output_files}. The arrays \texttt{write\_*[]} determine whether the values for each particle are written or not, thus potentially saving disk memory. Results are converted from GeV to CGS units.
	
	\item \texttt{double contribution\_instantaneous(int j, int counter, int k, double\\ **instantaneous\_primary\_spectra, double ****tables, double *initial\_energies, double\\ *final\_energies, int particle\_type, int hadronization\_choice)}\newline
	This routine computes one part $i$ of the instantaneous integrand of Eq.~\eqref{eq:secondary} (in GeV$^{-1}$ $\rightarrow$ GeV$^{-2}\cdot$s$^{-1}$) for the secondary particle \texttt{particle\_type}, initial energy $E^\prime =$ \texttt{energies[j]}, corresponding tabulated initial energy \texttt{initial\_energies[counter]} and final energy $E =$ \texttt{final\_energies[k]}. The sum over channels of production of the secondary particles may depend on the structure of the hadronization tables.
	
	\item \texttt{void hadronize\_instantaneous(double ***instantaneous\_hadronized\_spectra, double ****tables,\\ double *initial\_energies, double *final\_energies, double **instantaneous\_primary\_spectra,\\ double *energies, struct param *parameters)}\newline
	This routine computes the instantaneous secondary Hawking spectra, that is to say the integrand of Eq.~\eqref{eq:secondary} for all secondary particles, all initial energies in \texttt{energies[]} and all final energies in \texttt{final\_energies[]}. It fills the array \texttt{instantaneous\_hadronized\_spectra[][][]} using the routine \texttt{contribution\_instantaneous}, with format [secondary particle][initial energy][final energy]. If the initial energy is not in the hadronization tables, the contribution is extrapolated (this is not the case for \texttt{hadronization\_choice = 3}).
	
	\item \texttt{void integrate\_initial\_energies\_instantaneous(double ***hadronized\_emission\_spectra, double\\ **integrated\_hadronized\_spectra, double *energies, double *final\_energies, struct param\\ *parameters)}\newline
	This routine computes the integral of Eq.~\eqref{eq:full_secondary} (dimensionless $\rightarrow$ GeV$^{-1}\cdot$s$^{-1}$) using the \texttt{trapeze} routine. The results are stored in the array \texttt{instantaneous\_integrated\_hadronized\_spectra[][]} with format [secondary particle][final energy].
	
	\item \texttt{void add\_*\_instantaneous(double **instantaneous\_primary\_spectra, double\\ **instantaneous\_integrated\_hadronized\_spectra, double *energies, double *final\_energies, \\struct param *parameters)}\newline
	These three routines add the contributions of the primary photons, neutrinos and electrons to the secondary produced ones. These (trivial) branching ratios are not included in the computation of the hadronization tables with \texttt{PYTHIA} or \texttt{HERWIG}. The value in term of final energies is interpolated in the primary spectrum and added to the hadronized spectrum \texttt{instantaneous\_integrated\_hadronized\_spectra[][]}.
	
	\item \texttt{void add\_FSR\_instantaneous(double **instantaneous\_primary\_spectra, double\\ **instantaneous\_integrated\_hadronized\_spectra, double *energies, double *final\_energies,\\ double *masses\_primary, struct param *parameters)}\newline
	This routine adds the final state radiation from charged primary particles (electrons, muons, pions) to the secondary photon spectrum in the case of \texttt{hadronization\_choice = 3}, see Appendix~\ref{app:hadronization_tables}.
	
	\item \texttt{void write\_instantaneous\_hadronized\_spectra(double\\ **instantaneous\_integrated\_hadronized\_spectra, double *hadronized\_energies, struct param\\ *parameters)}\newline
	This routine writes the instantaneous secondary Hawking spectra in the file:
	
	\citecode{results/destination\_folder/instantaneous\_secondary\_spectra.txt}
	
	detailed in Section~\ref{sec:output_files}. The results are converted from GeV to CGS units.
\end{itemize}

\subsection{Auxiliary routines}

Fourteen auxiliary routines are used throughout the code:
\begin{itemize}
	\item \texttt{double trapeze(double x1, double x2, double y1, double y2)}\newline
	This routine performs the trapeze integration of a function $f$ that takes values \texttt{y1} in \texttt{x1} and \texttt{y2} in \texttt{x2} through
	\begin{equation}
		\int_\texttt{x1}^\texttt{x2} f(x){\rm d}x \approx \dfrac{1}{2}(\texttt{x2} - \texttt{x1})\times(\texttt{y1} + \texttt{y2})\,.
	\end{equation}
	
	\item \texttt{void free1D\_double(double *array)}, \texttt{void free1D\_int(int *array)}, \texttt{void free2D\_int(int **array, int l\_1stD)}, \texttt{void free2D\_double(double **array, int l\_1stD)}, \texttt{void free2D\_char(char **array, int l\_1stD)}, \texttt{void free3D\_double(double ***array, int l\_1stD, int l\_2ndD)}, \texttt{void free3D\_int(int\\ ***array, int l\_1stD, int l\_2ndD)} and	\texttt{void free4D\_double(double\\ ****array, int l\_1stD, int l\_2ndD, int l\_3rdD)}\newline
	These routines perform a proper memory freeing of $n-$dimensional arrays of various types, by recursively applying the native \texttt{free} routine.
	
	\item \texttt{int ind\_max(double *table, int llength)}\newline
	This routine returns the index of the maximum of the array \texttt{table[]} of length \texttt{llength}.
	
	\item \texttt{void fusion(double *table, int start1, int end1, int end2)}, \texttt{void sort\_fusion\_bis(double *table, int start, int end)} and \texttt{void sort\_fusion(double *table, int llength)}\newline
	These 3 routines perform a fusion sorting of the table \texttt{table} of length \texttt{llength}.
\end{itemize}

\subsection{General feature}

We have defined arrays \texttt{compute[]} that contain either 0's or 1's for all the primary and secondary particles and decide whether the contribution of the corresponding particle will be taken into account in the computation of the Hawking spectra. For example, if a ``0'' is set for the primary photon computation, then its primary spectrum will be all 0's and thus will not contribute to the secondary spectra. If a ``0'' is set for the secondary photon computation, then the secondary particles generated by primary photons are not taken into account in the secondary spectra (but the primary spectra of photons is still computed). These arrays are by default all 1's and should be modified only if the user wants to test specific scenarios or correct a bug.


\section{BlackHawk programs}
\label{sec:programs}

The \texttt{BlackHawk} code is split into two programs, which are presented in this Section:
\begin{itemize}
	\item \texttt{BlackHawk\_tot}: full time-dependent Hawking spectra,
	\item \texttt{BlackHawk\_inst}: instantaneous Hawking spectra.
\end{itemize}
Once a set of parameters is chosen, the two programs can be launched in the same \texttt{destination\_folder/} because the output files will not enter in conflict. We will now describe the structure of the \texttt{main} routines together with screen output examples.

\subsection{Common features}
\label{subsec:common_features}

When running the \texttt{BlackHawk} code, some routines will be called regardless of the program choice. First, some general quantities are fixed (which are converted into GeV when applicable, see Appendix~\ref{app:units}):
\begin{itemize}
	\item \texttt{machine\_precision} $= 10^{-10}$ defines the precision up to which two \texttt{double} numbers are considered as equal,
	\item \texttt{G} $= 6.67408\times 10^{-11}\,$m$^3\cdot$kg$^{-1}\cdot$s$^{-2}$ is the Newton constant in SI units,
	\item \texttt{Mp} $\equiv G^{-1/2}$ is the Planck mass in the natural system of units,
	\item \texttt{m\_*} are the masses of the SM elementary and composite particles (see Tables~\ref{tab:particles}, \ref{tab:BBN_particles} and \ref{tab:today_particles} in Appendix~\ref{app:particle_info}),
	\item \texttt{*\_conversion} are the quantities used to convert units from CGS/SI to GeV (see Appendix~\ref{app:units}).
\end{itemize}
The code further works in several steps, which are separated on the output screen. A new step starts with:

\citecode{[main] : ***** ...}

\noindent and ends with:

\citecode{DONE}

\noindent If the \texttt{full\_output} parameter is set to 1, more information will be displayed about the progress of the steps and an interactive pre-run check will be performed (overwrite existing data? check for memory usage?). In the case where information appears with the name of another routine inside brackets, it means that an error occurred, or that a warning must be displayed due to an unusual use of the code.

The first common step is the definition and filling of the parameters structure using \texttt{read\_params}. If \texttt{full\_output} is set to 1, an estimation of the memory that will be used is displayed by \texttt{memory\_estimation}. The user can choose to go on or to cancel the run. If no error is found in the input parameters, the output directory:

\citecode{results/destination\_folder/}

\noindent is created. If it already exists and if \texttt{full\_output} is set to 1, the user has the choice to overwrite the existing data or to stop the execution in order to choose another output folder, otherwise data will be overwritten automatically. For a subsequent data interpretation, the parameters file is copied in the output folder:

\citecode{results/destination\_folder/parameters.txt}

\noindent The expected output at this stage if of the form (\texttt{full\_output = 1}, program \texttt{BlackHawk\_inst}):\newline \newline
\texttt{
	\indent\hspace{1.45cm}\#\#\#\#\#\#\#\#\#\#\#\#\#\#\#\#\#\#\#\#\#\#\#\#\#\#\#\#\newline
	\indent\hspace{1.45cm}\#\hspace{1.12cm}BLACKHAWK v2.0\hspace{1.12cm}\#\newline
	\indent\hspace{1.45cm}\#\hspace{0.94cm}HAWKING SPECTRUM\hspace{0.94cm}\#\newline
	\indent\hspace{1.45cm}\#\hspace{0.76cm}COMPUTATION DEVICE\hspace{0.76cm}\#\newline
	\indent\hspace{1.45cm}\#\#\#\#\#\#\#\#\#\#\#\#\#\#\#\#\#\#\#\#\#\#\#\#\#\#\#\#\newline \newline
	[main] : STARTING EXECUTION...\newline
	[main] : READING THE RUN PARAMETERS IN 'parameters.txt'...  \hfill DONE\newline
	[main] : ESTIMATION OF THE MEMORY USE...\newline \newline
	\indent\hspace{1.55cm}Running this session will use at least 121.447 MB of RAM and 0.068 MB of disc memory.\newline \newline
	\indent\hspace{1.55cm}Do you want to continue? (type y or n) y\newline \newline
	\indent\hspace{1.55cm}\hfill DONE\newline
	[main] : SAVING RUN PARAMETERS...       \hfill DONE
}\newline

\noindent The subsequent execution steps depend on the program. In the following, screen output examples are given in the mode \texttt{full\_output = 0}. The results associated with the built-in \texttt{parameters.txt} file are given in Appendix~\ref{app:example_results} and should be used as checks at the first \texttt{BlackHawk} use.

\subsection{BlackHawk\_tot: Full time-dependent Hawking spectra}
\label{session0}

In this program, \texttt{BlackHawk} computes the time-dependent Hawking spectra of a chosen initial distribution of BHs, that is the quantities $\d^2 n_i/\d t\d E$ at all time intervals $t$ from the initial time $t\mrm{min}$ to the end of evaporation.

\texttt{BlackHawk} computes the initial distribution of BHs at $t\mrm{min}$ using the routine \texttt{spectrum} or reads the user-defined BH distribution file \texttt{table} with the routine \texttt{read\_users\_table} (depending on the \texttt{spectrum\_choice}), filling the arrays \texttt{init\_masses[]}, \texttt{init\_params[]} and \texttt{spec\_table[][]}. It writes the results in the output with \texttt{write\_spectrum}.

It then reads the relevant $f(M,x_j)$ (and $g(M,a^*)$ for Kerr BHs) tables, depending on the \texttt{metric} choice, using the \texttt{read\_fM\_table} (and \texttt{read\_gM\_table}) routines, filling the arrays \texttt{fM\_table[][]}, \texttt{gM\_table[][]}, \texttt{fM\_masses[]} and \texttt{fM\_param[]}, in order to evolve in time each initial BH mass and secondary parameter down to the Planck or remnant mass using the routine \texttt{life\_evolution}. This fills the arrays \texttt{life\_times[]}, \texttt{life\_masses[][][]}, \texttt{life\_params[][][]} and \texttt{dts[]}. The evolutions in time are written in the output using the routine \texttt{write\_life\_evolutions}.

Then \texttt{BlackHawk} reads the relevant greybody factor tables with the \texttt{read\_gamma\_tables} routine, depending on the \texttt{metric} choice, filling the arrays \texttt{gammas[][][]}, \texttt{gamma\_param[]} and \texttt{gamma\_x[]}, and the fit tables using \texttt{read\_asymp\_fits}, filling the array \texttt{fits[][][]}.

If the parameter \texttt{primary\_only} has been set to 0, \texttt{BlackHawk} reads the suitable hadronization tables (depending on the \texttt{hadronization\_choice}) with the routine \texttt{read\_hadronization\_tables}, filling the arrays \texttt{tables[][][][]}, \texttt{initial\_energies[]} and \texttt{final\_energies[]}. It uses these tables to compute the primary and secondary (if \texttt{primary\_only = 0}) Hawking spectra using the routine \texttt{total\_spectra}. Due to the large number of intermediate timesteps when a full distribution is considered, we do not perform the full computation in one step in the RAM memory, but rather do it timestep after timestep using the intermediate arrays \texttt{partial\_primary\_spectra[][]}, \texttt{partial\_hadronized\_spectra[][][]} and \texttt{partial\_integrated\_hadronized\_spectra[][]}, and the instantaneous routines \texttt{hadronize\_instantaneous}, \texttt{integrate\_initial\_energies\_instantaneous} and \texttt{add\_*\_instantaneous}. The intermediate results are written in the output using \texttt{write\_lines}.

This is the end of the execution of \texttt{BlackHawk\_tot}. The expected output is of the form:\newline \newline
\texttt{
	[main] : COMPUTING THE INITIAL DISTRIBUTION OF BLACK HOLES... \hfill DONE\newline
	[main] : WRITING INTO FILE 'BH\_spectrum.txt'...             \hfill     DONE\newline
	[main] : READING EVOLUTION TABLES...                          \hfill   DONE\newline
	[main] : COMPUTING THE EVOLUTION OF BLACK HOLES...            \hfill   DONE\newline
	[main] : WRITING INTO FILE 'life\_evolutions.txt'...          \hfill    DONE\newline
	[main] : READING GAMMA TABLES...                               \hfill  DONE\newline
	[main] : READING FIT TABLES...                           \hfill       DONE\newline
	[main] : READING HADRONIZATION TABLES...                \hfill         DONE\newline
	[main] : COMPUTING SPECTRA...                            \hfill    DONE\newline
	[main] : END OF EXECUTION
}

\subsection{BlackHawk\_inst: Instantaneous Hawking spectra}
\label{session1}

In this program, \texttt{BlackHawk} computes the instantaneous Hawking spectra of a distribution of BHs. That is, the quantities $\d^2 n_i/\d t\d E$ at the initial masses and secondary parameters, without time evolution.

First \texttt{BlackHawk} computes the initial distribution of BHs using the routine \texttt{spectrum} or it reads the user-defined BH distribution file \texttt{table} with the routine \texttt{read\_users\_table} (depending on the \texttt{spectrum\_choice}), filling the arrays \texttt{init\_masses[]}, \texttt{init\_params[]} and \texttt{spec\_table[][]}. It writes the results in the output with \texttt{write\_spectrum}.

Then \texttt{BlackHawk} reads the relevant greybody factor tables (depending on the \texttt{metric} choice) using the routine \texttt{read\_gamma\_tables}, filling the arrays \texttt{gammas[][][]}, \texttt{gamma\_masses[]} and \texttt{gamma\_x[]} and the fit table with the routine \texttt{read\_asymp\_fits}, filling the array \texttt{fits[][][]}. It computes the primary Hawking spectra using the routine \texttt{instantaneous\_primary\_spectrum}, filling the arrays \texttt{instantaneous\_primary\_spectra[][]}. The results are written in the output by \texttt{write\_instantaneous\_primary\_spectra}.

If the parameter \texttt{primary\_only} has been set to 0, \texttt{BlackHawk} reads the relevant hadronization tables (depending on the \texttt{hadronization\_choice}) using the routine \texttt{read\_hadronization\_tables}, filling the arrays \texttt{tables[][][][]}, \texttt{initial\_energies[]} and \texttt{final\_energies[]}, and uses them to compute the secondary Hawking spectra using the routine \texttt{hadronize\_instantaneous}, filling the array \texttt{instantaneous\_hadronized\_spectra[][][]}.

The initial energy dependence of the spectra is integrated out with \texttt{integrate\_initial\_energies\_instantaneous}, which fills the array \texttt{instantaneous\_integrated\_hadronized\_spectra[][]}. The contributions from primary photons, neutrinos and electrons are added to the secondary spectra by the routines \texttt{add\_*\_instantaneous}. The results are written in the output by the routine \texttt{write\_instantaneous\_hadronized\_spectra}.

This is the end of the execution of \texttt{BlackHawk\_inst}. The expected output is of the form:\newline \newline
\texttt{
	[main] : COMPUTING THE INITIAL DISTRIBUTION OF BLACK HOLES...          \hfill  DONE\newline
	[main] : WRITING INTO FILE 'BH\_spectrum.txt'...                      \hfill    DONE\newline
	[main] : READING GAMMA TABLES...                                \hfill         DONE\newline
	[main] : READING FIT TABLES...                               \hfill           DONE\newline
	[main] : COMPUTING PRIMARY SPECTRA...                              \hfill      DONE\newline
	[main] : WRITING INTO FILE 'instantaneous\_primary\_spectra.txt'...\hfill DONE\newline
	[main] : READING HADRONIZATION TABLES...                          \hfill       DONE\newline
	[main] : HADRONIZING PARTICLES...                                    \hfill    DONE\newline
	[main] : INTEGRATING OVER INITIAL ENERGIES...                     \hfill       DONE\newline
	[main] : WRITING INTO FILE 'instantaneous\_secondary\_spectra.txt'...\hfill DONE\newline
	[main] : END OF EXECUTION
}


\section{Output files}
\label{sec:output_files}

All the output files generated by a run of \texttt{BlackHawk} will be stored in the folder:

\citecode{results/destination\_folder/}

\noindent In this Section we describe the format of the files created by each program. Examples of results can be found in Appendix~\ref{app:example_results}. In all cases, the parameter file \texttt{parameters.txt} used for the run is copied in the output folder in order to allow for subsequent data interpretation. \texttt{Python} vizualisation scripts have been provided in the folder:

\citecode{scripts/visualization\_scripts/*.py}

\noindent in order to plot the data produced by both programs. They come with a \texttt{README} that explains how to configure them. The user can of course modify these scripts or use any other plotting program.

\subsection{BlackHawk\_tot program}

Running \texttt{BlackHawk\_tot} produces five (or four if \texttt{primary\_only} is set to 1) types of output files:
\begin{itemize}
	\item \texttt{BH\_spectrum.txt}: this file is written by the routine \texttt{write\_spectrum}. It contains the initial density spectrum of BHs and is 2-dimensional: the first column is a list of the BH initial masses (in g), the first line a list of the initial secondary parameters (dimensionless) and the bulk comoving number densities (in cm$^{-3}$).
	
	\item \texttt{life\_evolutions.txt}: this file is written by \texttt{write\_life\_evolutions}. It contains all the integrated timesteps for each initial BH mass and secondary parameter. It includes the total number of integration timesteps in the description line. It also contains \texttt{BH\_number} tables in which the first column gives the time (in s), and each other pair of columns is the evolution of the mass (in g) and secondary parameter (dimensionless) of BHs as a function of time for a fixed initial mass.
	
	\item \texttt{dts.txt}: this file is also written by \texttt{write\_life\_evolutions}. It contains the integrated timesteps $t$ (first column) and the corresponding time intervals $\d t$ (second column), both in s.
	
	\item \texttt{*\_primary\_spectrum.txt}: these files are written by the routine \texttt{write\_lines}. They contain the emission rates of each primary particle at each final time and for each simulated initial energy. The first line gives the list of energies (in GeV), the first column gives the list of times (in s), and each further column is the emission rate $\d^2 n_i/\d t\d E$ of Eq.~\eqref{eq:full_primary} of the particle per unit energy, time and covolume (in GeV$^{-1}\cdot$s$^{-1}\cdot$cm$^{-3}$).
	
	\item \texttt{*\_secondary\_spectrum.txt}: these files are also written by \texttt{write\_lines}. They contain the emission rates of each secondary particles at each final times and for each simulated final energies. The first line gives the list of energies (in GeV), the first column gives the list of times (in s), and each other column is the emission rate $\d^2 n_j/\d t\d E$ of Eq.~\eqref{eq:full_secondary} of the particle per units of energy, time and covolume (in GeV$^{-1}\cdot$s$^{-1}\cdot$cm$^{-3}$). These files will not be generated if the parameter \texttt{primary\_only} has been set to 1.
\end{itemize}

\subsection{BlackHawk\_inst program}

Running \texttt{BlackHawk\_inst} produces three (or two if \texttt{primary\_only} is set to 1) output files:
\begin{itemize}
	\item \texttt{BH\_spectrum.txt}: this file is the same as for \texttt{BlackHawk\_tot}.
	
	\item \texttt{instantaneous\_primary\_spectra.txt}: this file is written by \texttt{write\_instantaneous\_primary\_spectra}. It contains the emission rates of the primary particles for each simulated initial energy. The first line is the list of primary particles, the first column is the list of energies (in GeV), and each other column is the emission rate $\d^2 n_i/\d t\d E$ of Eq.~\eqref{eq:full_primary} per unit energy and time (in GeV$^{-1}\cdot$s$^{-1}\cdot$cm$^{-3}$).
	
	\item \texttt{instantaneous\_secondary\_spectra.txt}: this file is written by \texttt{write\_instantaneous\_hadronized\_spectra}. It contains the emission rates of the secondary particles for each simulated final energy. The first line is the list of secondary particles, the first column is that of energies, and each other column is the emission rate $\d^2 n_j/\d t\d E$ of Eq.~\eqref{eq:full_secondary} per unit energy and time (in GeV$^{-1}\cdot$s$^{-1}\cdot$cm$^{-3}$). It will not be generated if the parameter \texttt{primary\_only} has been set to 1.
\end{itemize}


\section{Memory usage}
\label{sec:memory_usage}

The code \texttt{BlackHawk} has been designed to minimize the memory used (both RAM and disk) and the computation time while avoiding excessive approximations. In this Section we give estimates of the memory used by each program.

\subsection{RAM used}

To every array defined in \texttt{BlackHawk}, a memory space is allocated with a \texttt{malloc} call. This memory is freed at the moment the array stops being necessary for the subsequent part of the run. Then, the RAM used by \texttt{BlackHawk} at a given step of a session can be estimated as a sum over all active arrays at that time. \texttt{double} numbers are coded in 8 bytes and \texttt{int} in 4 bytes. Memory spaces $\mathcal{M}$ are given in bytes. We assume that $\meanN \sim 1000$ timesteps are necessary for the integration of one BH mass and secondary parameter. For \texttt{BlackHawk\_tot} we have:
\begin{itemize}
	\item step 1 (BH spectrum):
	\begin{itemize}
		\item $\texttt{init\_masses[]} = 8\times\texttt{BH\_number}$,
		\item $\texttt{init\_params[]} = 8\times\texttt{BH\_number}$,
		\item $\texttt{spec\_table[]} = 8\times\texttt{BH\_number}\times\texttt{param\_number}$,
	\end{itemize}
	\item step 2 (BH evolution):
	\begin{itemize}
		\item $\texttt{init\_masses[]} = 8\times\texttt{BH\_number}$,
		\item $\texttt{init\_params[]} = 8\times\texttt{BH\_number}$,
		\item $\texttt{spec\_table[]} = 8\times\texttt{BH\_number}\times\texttt{param\_number}$,
		\item $\texttt{fM\_table[][]} = 8\times\texttt{nb\_fM\_param}\times\texttt{nb\_fM\_masses}$,
		\item $\texttt{gM\_table[][]} = 8\times\texttt{nb\_fM\_param}\times\texttt{nb\_fM\_masses}$ if \texttt{metric = 0},
		\item $\texttt{fM\_masses[]} = 8\times\texttt{nb\_fM\_masses}$,
		\item $\texttt{fM\_param[]} = 8\times\texttt{nb\_fM\_param}$,
		\item $\texttt{life\_masses[][][]} = 8\times\texttt{BH\_number}^2\times\texttt{param\_number}^2\times\texttt{limit}$,
		\item $\texttt{life\_params[][][]} = 8\times\texttt{BH\_number}^2\times\texttt{param\_number}^2\times\texttt{limit}$,
		\item $\texttt{life\_times[]} = 8\times\texttt{BH\_number}\times\texttt{param\_number}\times\texttt{limit}$,
		\item $\texttt{dts[]} = 8\times\texttt{BH\_number}\times\texttt{param\_number}\times\texttt{limit}$,
	\end{itemize}
	\item step 3 (primary and secondary spectra):
	\begin{itemize}
		\item $\texttt{spec\_table[]} = 8\times\texttt{BH\_number}\times\texttt{param\_number}$,
		\item $\texttt{life\_masses[][][]} = 8\times\texttt{BH\_number}^2\times\texttt{param\_number}^2\times\texttt{limit}$,
		\item $\texttt{life\_params[][][]} = 8\times\texttt{BH\_number}^2\times\texttt{param\_number}^2\times\texttt{limit}$,
		\item $\texttt{life\_times[]} = 8\times\texttt{BH\_number}\times\texttt{param\_number}\times\texttt{limit}$,
		\item $\texttt{dts[]} = 8\times\texttt{BH\_number}\times\texttt{param\_number}\times\texttt{limit}$,
		\item $\texttt{gammas[][][]} = 8\times4\times \texttt{nb\_gamma\_param}\times\texttt{nb\_gamma\_x}$,
		\item $\texttt{gamma\_param[]} = 8\times\texttt{nb\_gamma\_param}$,
		\item $\texttt{gamma\_x[]} = 8\times\texttt{nb\_gamma\_x}$,
		\item $\texttt{fits[][][]} = 8\times4\times\texttt{nb\_gamma\_param}\times \texttt{nb\_gamma\_fits}$,
		\item $\texttt{dof[]} = 8\times(\texttt{particle\_number} + \texttt{grav} + \texttt{add\_DM})$,
		\item $\texttt{spins[]} = 8\times(\texttt{particle\_number} + \texttt{grav} + \texttt{add\_DM})$,
		\item $\texttt{masses\_primary[]} = 8\times(\texttt{particle\_number} + \texttt{grav} + \texttt{add\_DM})$,
		\item $\texttt{times[]} \approx 8\times \meanN \times\texttt{BH\_number}\times\texttt{param\_number}$,
		\item $\texttt{energies[]} = 8\times\texttt{E\_number}$,
		\item $\texttt{tables[][][][]} = 8\times\texttt{nb\_fin\_part}\times\texttt{nb\_init\_en}\times\texttt{nb\_fin\_en}\times\texttt{nb\_fin\_part}$,
		\item $\texttt{initial\_energies[]} = 8\times\texttt{nb\_init\_en}$,
		\item $\texttt{final\_energies[]} = 8\times\texttt{nb\_fin\_en}$,
		\item $\texttt{partial\_hadronized\_spectra[][][]} = 8\times\texttt{nb\_fin\_part}\times\texttt{E\_number}\times\texttt{nb\_fin\_en}$,
		\item $\texttt{partial\_primary\_spectra[][]} = 8\times(\texttt{particle\_number} + \texttt{grav} + \texttt{add\_DM})\times\texttt{E\_number}$,
		\item $\texttt{partial\_integrated\_hadronized\_spectra[][]} = 8\times\texttt{nb\_fin\_part}\times\texttt{nb\_fin\_en}$,
		\item $\texttt{masses\_secondary[]} = 8\times\texttt{nb\_fin\_part}$.
	\end{itemize}
\end{itemize}
Using the parameters of Appendix~\ref{app:example_parameters}, the arrays occupy at most $\mathcal{M}\sim120\,$MB, mostly because of the hadronization tables.

For \texttt{BlackHawk\_inst} we have:
\begin{itemize}
	\item step 1 (BH spectrum):
	\begin{itemize}
		\item $\texttt{BH\_masses[]} = 8\times\texttt{BH\_number}$,
		\item $\texttt{BH\_params[]} = 8\times\texttt{param\_number}$,
		\item $\texttt{spec\_table[][]} = 8\times\texttt{BH\_number}\times\texttt{param\_number}$,
	\end{itemize}
	\item step 2 (primary spectra):
	\begin{itemize}
		\item $\texttt{BH\_masses[]} = 8 \times \texttt{BH\_number}$,
		\item $\texttt{BH\_params[]} = 8 \times \texttt{param\_number}$,
		\item $\texttt{spec\_table[][]} = 8 \times \texttt{BH\_number}\times\texttt{param\_number}$,
		\item $\texttt{gammas[][][]} = 8 \times 4 \times  \texttt{nb\_gamma\_param} \times \texttt{nb\_gamma\_x}$,
		\item $\texttt{gamma\_param[]} = 8 \times \texttt{nb\_gamma\_param}$,
		\item $\texttt{gamma\_x[]} = 8 \times \texttt{nb\_gamma\_x}$,
		\item $\texttt{fits[][][]} = 8 \times 4 \times \texttt{nb\_gamma\_param} \times \texttt{nb\_gamma\_fits}$,
		\item $\texttt{dof[]} = 8 \times (\texttt{particle\_number} + \texttt{grav} + \texttt{add\_DM})$,
		\item $\texttt{spins[]} = 8 \times (\texttt{particle\_number} + \texttt{grav} + \texttt{add\_DM})$,
		\item $\texttt{masses\_primary[]} = 8 \times (\texttt{particle\_number} + \texttt{grav} + \texttt{add\_DM})$,
		\item $\texttt{instantaneous\_primary\_spectra[][]} = 8 \times (\texttt{particle\_number} + \texttt{grav} + \texttt{add\_DM}) \times \texttt{E\_number}$,
		\item $\texttt{energies[]} = 8 \times\texttt{E\_number}$,
	\end{itemize}
	\item step 3 (during hadronization):
	\begin{itemize}
		\item $\texttt{instantaneous\_primary\_spectra[][]} = 8 \times (\texttt{particle\_number} + \texttt{grav} + \texttt{add\_DM}) \times \texttt{E\_number}$,
		\item $\texttt{energies[]} = 8 \times \texttt{E\_number}$,
		\item $\texttt{tables[][][][]} = 8 \times \texttt{nb\_fin\_part} \times \texttt{nb\_init\_en} \times \texttt{nb\_fin\_en} \times \texttt{nb\_fin\_part}$,
		\item $\texttt{initial\_energies[]} = 8 \times \texttt{nb\_init\_en}$,
		\item $\texttt{final\_energies[]} = 8 \times \texttt{nb\_fin\_en}$,
		\item $\texttt{masses\_secondary[]} = 8 \times \texttt{nb\_fin\_part}$,
		\item $\texttt{instantaneous\_hadronized\_spectra[][][]} = 8 \times \texttt{nb\_fin\_part} \times \texttt{E\_number} \times \texttt{nb\_fin\_en}$,
	\end{itemize}
	\item step 3 bis (during integration):
	\begin{itemize}
		\item $\texttt{instantaneous\_primary\_spectra[][]} = 8 \times (\texttt{particle\_number} + \texttt{grav} + \texttt{add\_DM}) \times \texttt{E\_number}$,
		\item $\texttt{energies[]} = 8 \times \texttt{E\_number}$,
		\item $\texttt{initial\_energies[]} = 8 \times \texttt{nb\_init\_en}$,
		\item $\texttt{final\_energies[]} = 8 \times \texttt{nb\_fin\_en}$,
		\item $\texttt{instantaneous\_hadronized\_spectra[][][]} = 8 \times \texttt{nb\_fin\_times} \times \texttt{E\_number} \times \texttt{nb\_fin\_en}$,
		\item $\texttt{instantaneous\_integrated\_hadronized\_spectra[][]} = 8 \times \texttt{nb\_fin\_part} \times \texttt{nb\_fin\_en}$.
	\end{itemize}
\end{itemize}
Using the parameters of Appendix~\ref{app:example_parameters}, the arrays occupy at most $\mathcal{M}\sim120\,$MB, mostly because of the hadronization tables.

\subsection{Static disk memory used}

The output generated is written in \texttt{.txt} files using a precision of 5 significant digits. Adding the exponent and the coma, we obtain 12 characters per written number, which is 12 bytes. For \texttt{BlackHawk\_tot} we have:
\begin{itemize}
	\item file \texttt{BH\_spectrum.txt}: $\mathcal{M} = 12 \times (\texttt{BH\_number}+1)\times(\texttt{param\_number}+1)$,
	\item file \texttt{life\_evolutions.txt}: $\mathcal{M}\approx 12 \times \texttt{BH\_number}\times(1. + 2\times\texttt{param\_number}) \times \texttt{BH\_number}\times\texttt{param\_number} \times \meanN$,
	\item files \texttt{*\_primary\_spectrum.txt}: $\mathcal{M} \approx 12 \times (\texttt{particle\_number} + \texttt{grav} + \texttt{add\_DM}) \times \texttt{E\_number} \times \meanN \times \texttt{BH\_number} \times\texttt{param\_number}$,
	\item files \texttt{*\_secondary\_spectrum.txt}: $\mathcal{M} \approx 12 \times \texttt{nb\_fin\_part} \times \texttt{nb\_fin\_en} \times \meanN \times \texttt{BH\_number}\times \texttt{param\_number}
	$.
\end{itemize}
Using the parameters of Appendix~\ref{app:example_parameters}, the total written disk space is $\mathcal{M}\sim 70\,$MB.

For \texttt{BlackHawk\_inst} we have:
\begin{itemize}
	\item file \texttt{BH\_spectrum.txt}: $\mathcal{M} = 12 \times 3 \times \texttt{BH\_number}$
	\item file \texttt{instantaneous\_primary\_spectra.txt}: $\mathcal{M} = 12 \times \texttt{E\_number} \times (\texttt{particle\_number} + \texttt{grav} + \texttt{add\_DM})$,
	\item file \texttt{instantaneous\_secondary\_spectra.txt}: $\mathcal{M} = 12 \times \texttt{nb\_fin\_en} \times\\ \texttt{nb\_fin\_part}$.
\end{itemize}
Using the parameters of Appendix~\ref{app:example_parameters}, the total written disk space is $\mathcal{M}\sim70\,$kB.


\section{Other applications}
\label{sec:other_applications}

In this Section we present some hints about how to modify \texttt{BlackHawk}. Most of these modifications will require adding extra parameters in the parameter files and thus a modification of the routines \texttt{read\_params} and \texttt{read\_*\_infos}, and of the structure \texttt{struct param}.

\subsection{Computing new numerical tables}
\label{subsec:tables}

The user may be interested in recomputing the tables described in Appendix~\ref{app:numerical_tables}, either to have more entries or to compute them with different methods for comparison. The easiest way to add tables in \texttt{BlackHawk} would be:
\begin{itemize}
	\item authorize the corresponding ``choice'' parameters to have other integer values,
	\item put the new tables in a new directory in the subfolder:
	
	\citecode{src/tables/}
	
	\item create/modify the \texttt{infos.txt} information file and the corresponding \texttt{read\_*\_infos} routine,
	\item add a \texttt{case} into the tables reading routines,
	\item make sure that the way tables are used in the routines will be compatible with the format of the new ones.
\end{itemize}
All the scripts used to compute the current tables are included in \texttt{BlackHawk} in the subfolder:

\citecode{scripts/}

\noindent together with \texttt{README} files.

\subsection{Using another BH mass and spin distribution}
\label{subsec:distribution}

The user may be interested in testing its own BH distribution. Here are the main steps to add a pre-built distribution:
\begin{itemize}
	\item add a ``choice'' parameter to the \texttt{struct param} choosing the distribution,
	\item add the corresponding analytical formula to the routines \texttt{M\_dist} and \texttt{param\_dist},
	\item modify the parameter \texttt{tmin}, or the built-in Eq.~\eqref{eq:formation_time} if the distribution is valid at a different initial time.
\end{itemize}
Providing a tabulated initial distribution to \texttt{BlackHawk} is done by switching the parameter \texttt{spectrum\_choice} to $-1$, putting the table file in the subfolder: 

\citecode{src/tables/users\_spectra/}

\noindent and giving its full file name (including the \texttt{.txt} extension for example) to the parameter \texttt{table}. The format has to be:
\begin{itemize}
	\item the first column for BH masses $M$, the first line for BH secondary parameter $x$ and the bulk of the table for the comoving number densities ${\rm d}n$ (with ${\rm d}M$ taken around $M$ and ${\rm d}x$ taken around $x$),
	\item masses and densities in CGS units (g and cm$^{-3}$ respectively), secondary parameters in dimensionless form,
	\item numbers in standard scientific notation,
	\item only additional text: a string of characters on the up-left corner (\textit{e.g.}~``mass/spin'' for Kerr BHs).
\end{itemize}

\subsection{Adding primary particles}
\label{subsec:primary}

If the user wants to add hypothetical primary Hawking particles, the following steps have to be undertaken:
\begin{itemize}
	\item increase the parameter \texttt{particle\_number} in the greybody factor table information file \texttt{infos.txt} or add the new particle(s) with a switch similar to the one for the graviton,
	\item recompute the $f(M,x_j)$ (and $g(M,a^*)$ for Kerr BHs) tables to account for this(ese) new emission(s),
	\item if the spin(s) of the new particle(s) is(are) not among the greybody factor tables, compute the new ones and modify the \texttt{read\_gamma\_tables} and \texttt{read\_gamma\_fits} routines,
	\item add the new particle(s) to all the fixed length arrays of particle types (for example the file names or columns in the writing routines),
	\item eventually add its(their) contribution(s) to the secondary spectra.
\end{itemize}
We want to point out that as the graviton emission, as well as DM emission, are included in \texttt{BlackHawk}, new layers of new particles may become difficult to handle.

\subsection{Adding secondary particles}
\label{subsec:secondary}

In order to add secondary Hawking particles to the code, one has to:
\begin{itemize}
	\item recompute the hadronization tables to take new branching ratios into account,
	\item add the new particle(s) to all the fixed length arrays of particle types (for example the file names or columns in the writing routines),
	\item add the corresponding contribution(s) to the routine \texttt{contribution\_instantaneous}.
\end{itemize}

\subsection{Other types of black holes}
\label{subsec:exotic_BH}

If the user wants to compute the Hawking emission of BHs different from the already included ones, several ingredients are needed:
\begin{itemize}
	\item add a switch to the parameter file to select amongst the new types of BHs,
	\item modify/add the Hawking temperature function \texttt{temp\_BH} for these BHs,
	\item modify/add evolution routines \texttt{loss\_rate\_*} and \texttt{life\_evolution},
	\item compute the corresponding $f$, $g$ and eventually new evolution parameters tables and add the corresponding reading routines and associated information in the \texttt{infos.txt} file,
	\item compute the new greybody factor tables and update the corresponding reading and interpolating routines \texttt{read\_gamma\_tables}, \texttt{read\_gamma\_fits} and \texttt{dNdtdE}.
\end{itemize}
Depending on the complexity of the BH model, the user may need to implement some or all of the above modifications. We want to point out that several BH metrics are already present inside \texttt{BlackHawk v2.0}: Reissner--Nordstr\"om BHs, higher-dimensional BHs and polymerized BHs, on top of the usual Kerr BHs.


\section{Conclusion}
\label{sec:conclusion}

\texttt{BlackHawk} is the first public code generating both primary and secondary Hawking evaporation spectra for any distribution of black holes, and their evolution in time.\footnote{While the latest version of the manual was being written, Refs.~\cite{Cheek:2021odj,Cheek:2021cfe} appeared, describing the code \texttt{ULYSSES} aimed at leptogenesis, in which Hawking radiation of primordial black holes is computed in a similar way as \texttt{BlackHawk}. Its authors have checked for consistency with the results from \texttt{BlackHawk}.} The primary spectra are obtained using greybody factors, and the secondary ones result from the decay and hadronization of the primary particles. The black holes and spectra evolution are obtained by considering the energy (and angular momentum for Kerr black holes) losses via Hawking radiation and the modification of the temperature of the black hole.
\texttt{BlackHawk} is designed in a user-friendly way and modifications can be easily implemented. The primary application is to study the effects of particles generated by Hawking evaporation on observable quantities and thus to disqualify or set constraints on cosmological models implying the formation of black holes, as well as to test the Hawking radiation assumptions and study black hole general properties.
\texttt{BlackHawk v2.0} embeds new features compared to the previous version, most of which deal with beyond Standard Model physics: emission of dark matter by Hawking radiation, spin $3/2$ particles or black hole solutions beyond standard general relativity and usual formation models (Reissner--Nordstr\"om, higher-dimensional and polymerized black holes). The code has thus been extended to allow for beyond the Standard Model studies (for illustrating examples and a full review of the \texttt{BlackHawk} related literature, see the release note~\cite{release_note}).

\section*{Acknowledgments}

We gratefully acknowledge helpful exchanges with P.~Richardson in particular on the hadronization procedure and the \texttt{HERWIG} code. We are also thankful to  J.~Silk for many constructive discussions, to P.~Skands for help with \texttt{PYTHIA} and hadronization, and to G.~Robbins for the interface with \texttt{SuperIso Relic}. We also thank L.~Morrison, S.~Profumo and A.~M.~Coogan for help using the \texttt{Hazma} package. The authors thank the CERN Theory Department for its hospitality during which part of this work was done.

\appendix


\section{Units}
\label{app:units}

The \texttt{BlackHawk} code uses the GeV unit internally in order to have simpler analytical expressions. However, to make the user interface more accessible, the input parameters as well as the output files are in CGS units. We provide below unit conversions from the natural system of units where $\hbar = c = k\mrm{B} = G = 4\pi\varepsilon_0 = 1$ to CGS or SI.

\subsection{Energy}

The energy conversion from GeV to Joules is
\begin{equation}
	E\mrm{J} = 1.602176565\times 10^{-10}\, E\mrm{GeV}\,.
\end{equation}

\subsection{Mass}

The dimensional link between energy and mass is $[m] = [E/c^2]$, and the conversion from GeV to grams is
\begin{equation}
	m\mrm{g} = 5.60958884\times 10^{23}\, m\mrm{GeV}\,.
\end{equation}

\subsection{Time}

The dimensional link between energy and time is $[t] = [\hbar/E]$, and the conversion from GeV to seconds is
\begin{equation}
	t\mrm{s} = 1.519267407\times 10^{24}\, t_{{\rm GeV}^{-1}}\,.
\end{equation}

\subsection{Distance}

The dimensional link between energy and distance is $[l] = [\hbar c / E]$, and the conversion from GeV to meters is
\begin{equation}
	l\mrm{cm} = 5.06773058\times 10^{13}\,l\mrm{{\rm GeV}^{-1}}\,.
\end{equation}

\subsection{Temperature}

The dimensional link between energy and temperature is $[T] = [E/k\mrm{B}]$, and the conversion from GeV to Kelvins is
\begin{equation}
	T\mrm{K} = 8.61733063\times 10^{-14}\,T\mrm{GeV}\,.
\end{equation}

\subsection{Electric charge}

The dimensional link between energy and electric charge is $[Q^2] = [4\pi\varepsilon_0\hbar c]$, hence electric charge is dimensionless in the natural system of units and we have, through the definition of the fine structure constant $\alpha$,
\begin{equation}
	\alpha \equiv \dfrac{e^2}{4\pi\varepsilon_0\hbar c}\,, \qquad \left.e\times q^*\right|\mrm{C} = \sqrt{\alpha}\times q^*\,,
\end{equation}
where $q^*$ is the charge number.


\section{Particle information}
\label{app:particle_info}
\newpage
\begin{table}[!ht]
	\centering{
		\begin{tabular}{|c|c|c|c|c|}
			\hline
			particle & symbol & mass (GeV/c$^2$) & spin & quantum dof $g_i$ \\
			\hline
			Higgs boson	  & $h^0$ & $1.2503\times 10^{2}$  & $0$ & $1$ \\
			photon 		  & $\gamma$  & $0$ 	  & $1$ & $2$ \\
			gluons 		  & $g$ 	  & $0$ 	  & $1$ & $16$ \\
			W bosons	  & $W^\pm$   & $8.0403\times 10^1$  & $1$ & $6$ \\
			Z boson		  & $Z^0$	  & $9.11876\times 10^1$ & $1$ & $3$ \\
			neutrinos	  & $\nu_{e,\mu,\tau},\overline{\nu}_{e,\mu,\tau}$ & $0$ & $1/2$ & $6$ \\
			electron	  & $e^{\pm}$ & $5.109989461\times 10^{-4}$ & $1/2$ & $4$ \\
			muon		  & $\mu^\pm$ & $1.056583745\times 10^{-1}$ & $1/2$ & $4$ \\
			tau			  & $\tau^\pm$ & $1.77686$ & $1/2$ & $4$ \\
			up quark 	  & $u,\overline{u}$ & $2.2\times 10^{-3}$ & $1/2$ & $12$ \\
			down quark 	  & $d,\overline{d}$ & $4.7\times 10^{-3}$ & $1/2$ & $12$ \\
			charm quark   & $c,\overline{c}$ & $1.27$ & $1/2$ & $12$ \\
			strange quark & $s,\overline{s}$ & $9.6\times 10^{-2}$ & $1/2$ & $12$ \\
			top quark 	  & $t,\overline{t}$ & $1.7321\times 10^2$ & $1/2$ & $12$ \\
			bottom quark  & $b,\overline{b}$ & $4.18$ & $1/2$ & $12$ \\
			graviton 	  & $G$ & $0$ & $2$ & $2$ \\
			\hline
		\end{tabular}
		\caption{Properties of the elementary particles of the Standard Model, in addition to the graviton~\cite{ParticleDataGroup:2018ovx}. The number of quantum dof is the product of the family, antiparticle, the colour and the helicity multiplicities. Neutrinos are here considered massless. In the code, gluons have been assigned an effective mass to account for the QCD energy scale $\Lambda \approx 200\,$MeV.\label{tab:particles}}
	}
\end{table}

\begin{table}[!ht]
	\centering{
		\begin{tabular}{|c|c|c|c|}
			\hline
			particle 			& symbol & mass (GeV/c$^2$)											& lifetime (s) \\
			\hline																	
			photon				& $\gamma$ & $0$											& $\infty$ \\
			electron			& $e^\pm$ & $5.109989461\times 10^{-4}$											& $\infty$ \\
			muon				& $\mu^\pm$	 & $1.056583745\times 10^{-1}$										& $(2.1969811\pm0.0000022)\times 10^{-6}$ \\
			neutrinos			& $\nu_{e,\mu,\tau},\overline{\nu}_{e,\mu,\tau}$	& 0 &  $\infty$ \\
			charged pions		& $\pi^\pm$	& $1.3957018\times 10^{-1}$										& $(2.6033\pm0.0005)\times 10^{-8}$ \\
			neutral ``long'' kaon	& $K^0\mrm{L}$ & $4.977\times 10^{-1}$										& $(5.099\pm0.021)\times 10^{-8}$ \\
			charged kaons		& $K^\pm$ & $4.937\times 10^{-1}$											& $(1.2379\pm0.0021)\times 10^{-8}$ \\
			proton				& p,$\overline{\rm p}$ & $9.38272\times 10^{-1}$								& $\infty$ \\
			neutron				& n,$\overline{\rm n}$ & $9.395654\times 10^{-1}$								& $880.2\pm 1$ \\
			\hline
		\end{tabular}
		\caption{Particles with a lifetime longer than $10^{-8}\,$s~\cite{ParticleDataGroup:2018ovx}, relevant for BBN studies and used to compute the hadronization tables in \texttt{pythia\_tables/} and \texttt{herwig\_tables/}.\label{tab:BBN_particles}}
	}
\end{table}

\begin{table}[!ht]
	\centering{
		\begin{tabular}{|c|c|c|c|}
			\hline
			particle 			& symbol & mass (GeV/c$^2$)											& lifetime (s) \\
			\hline																	
			photon				& $\gamma$ & 0											& $\infty$ \\
			electron			& $e^\pm$ & $5.109989461\times 10^{-4}$											& $\infty$ \\
			neutrinos			& $\nu_{e,\mu,\tau},\overline{\nu}_{e,\mu,\tau}$	& 0 & $\infty$ \\
			proton				& p,$\overline{\rm p}$ & $9.38272\times 10^{-1}$								& $\infty$ \\
			\hline
		\end{tabular}
		\caption{Stable particles~\cite{ParticleDataGroup:2018ovx}, relevant for evaporating BHs in the low redshift universe and used to compute the hadronization table in \texttt{pythia\_tables\_new/} and \texttt{hazma\_tables/}.\label{tab:today_particles}}
	}
\end{table}


\section{Numerical tables}
\label{app:numerical_tables}

In this Appendix, we detail the way the numerical tables are computed inside \texttt{BlackHawk}. While increasing the transparency of the code, this also gives hints concerning how to modify the tables or compute them with different methods.

\subsection{Greybody factors}
\label{app:greybody_factors}

\subsubsection{General framework}

The greybody factors describe the probability that a particle, generated by quantum fluctuations at the horizon of a BH, escapes to spatial infinity. They are the only factor differing from the blackbody radiation in Eq.~\eqref{eq:hawking_master}. They are obtained by solving the equations of motion of (beyond the) SM particles in curved spacetime, with special boundary conditions (for a detailed discussion on quantum fields, see \textit{e.g.}~\cite{Weinberg:1995mt,Weinberg:1996kr,Weinberg:2000cr}). For massive bosons and spin $1/2$ fermions, the free equations of motion are respectively the Proca and Dirac equations
\begin{equation}
	(\square + \mu^2)\Psi = 0\,, \qquad (\gamma^\mu\partial_\mu - i\mu)\Psi = 0\,,
\end{equation}
while for massive spin $3/2$ fermions it is the Rarita--Schwinger equation
\begin{equation}
	(\epsilon^{\mu\kappa\rho\nu} \gamma_5\gamma_\kappa\partial_\rho - i\mu\sigma^{\mu\nu})\Psi_\nu = 0\,,
\end{equation}
where $\mu$ is the particle rest mass and $\gamma_\mu$ are the Dirac matrices, with $\sigma^{\mu\nu} \equiv i[\gamma^\mu,\gamma^\nu]/2$. Then, we proceed to a separation of these equations into a radial and an angular part; the separation is always possible for spherically symmetric and static BHs, as well as for Kerr BHs. To do that, we decompose the wave function $\Psi$ using the symmetries of the metric (Killing vectors $\partial_t$ and $\partial_\varphi$) in the Boyer--Lindquist coordinates $t,r,\theta,\varphi$
\begin{equation}
	\Psi(t,r,\theta,\varphi) = e^{-i E t}\psi(r)S_{lm}^{s}(\theta,\varphi)\,,
\end{equation}
where $E$ is the particle energy and $S_{lm}^s$ is the spin-weighted spheroidal harmonics for angular momentum $l$, projection $m$ and field spin $s$. The angular part of the wave equation gives some separation constant $\lambda_{s,lm}$ whose analytical expansion in terms of $\gamma \equiv EMa^*$ for Kerr BHs is given \textit{e.g.}~in~\cite{Dong:2015yjs} (valid for all spherically symmetric and static BHs as well with $a^* = 0$), and the radial part satisfies what is denoted as the Teukolsky equation, whose general mathematical form is very complicated (for a complete mathematical overview of this topic, see \textit{e.g.}~\cite{Chandrasekhar:1985kt}). This equation can be numerically solved directly, but the convergence of the solutions at the integration boundaries --- namely at BH horizon and at spatial infinity --- is not evident. Several authors have thus further transformed this equation into a Schr\"odinger-like wave equation with short-ranged potentials, different for each field spin, allowing for a more careful numerical treatment.

\subsubsection{Kerr black holes}

Chandrasekhar and Detweiler have shown that the Teukolsky equation can be reduced to a Schr\"odinger-like wave equation for Kerr BHs~\cite{chandra1,chandra2,chandra3,chandra4}. They found necessary to define a Eddington--Finkelstein radial coordinate $r^*$
\begin{equation}
	\dfrac{{\rm d}r^*}{{\rm d}r} = \dfrac{\rho^2}{\Delta}\,, \label{eq:tortoise}
\end{equation}
where $\rho(r)^2 \equiv r^2 + \alpha^2$ and $\alpha^2 \equiv a^2+am/E$, $a$ being the BH spin and $m$ the projection of the angular momentum $l$. This equation can be integrated to give
\begin{equation}
	r^*(r) = r + \dfrac{r\mrm{S}r_+ + am/E}{r_+ - r_-}\ln\left( \dfrac{r}{r_+} - 1 \right) - \dfrac{r\mrm{S}r_- + am/E}{r_+ - r_-}\ln\left( \dfrac{r}{r_-} - 1 \right),
\end{equation}
where $r\mrm{S} = 2M$ is the BH Schwarzschild radius. Unfortunately, the inverse of this equation has to be found numerically and is generally difficult to determine with accurate precision. The Schr\"odinger-like wave equation is for all spins $s$
\begin{equation}
	\dfrac{{\rm d}^2\psi_s}{{\rm d}r^{*2}} + \left( E^2 - V_{s}(r^*) \right)\psi_s = 0\,. \label{eq:kerrRW}
\end{equation}
The method to transform the Teukolsky equation into this simple wave equation was proposed in the Chandrasekhar and Detweiler papers~\cite{chandra1,chandra2,chandra3,chandra4}. It is indeed difficult to find short-range potentials allowing for precise numerical computations. They give the form of such potentials in \cite{chandra1,chandra2} for spin 2, \cite{chandra3} for spins 0 and 1 and \cite{chandra4} for spin $1/2$. The potentials are\footnote{We found that the spin 0 potential had a missing ``$r$'' in \cite{chandra3}.}
\begin{align}
	&V_0(r) = \dfrac{\Delta}{\rho^4}\left( \lambda_{0\,lm} + \dfrac{\Delta + 2r(r-M)}{\rho^2} - \dfrac{3r^2\Delta}{\rho^4} \right), \\
	&V_{1/2,\pm}(r) = (\lambda_{1/2\,lm}+1)\dfrac{\Delta}{\rho^4} \mp \dfrac{\sqrt{(\lambda_{1/2,l,m}+1)\Delta}}{\rho^4}\left( (r-M) - \dfrac{2r\Delta}{\rho^2} \right), \\
	&V_{1,\pm}(r) = \dfrac{\Delta}{\rho^4}\left( (\lambda_{1\,lm}+2)-\alpha^2\dfrac{\Delta}{\rho^4} \mp i\alpha\rho^2 \dfrac{{\rm d}}{{\rm d}r}\left( \dfrac{\Delta}{\rho^4} \right) \right), \\
	&V_{2}(r) = \dfrac{\Delta}{\rho^8}\bigg( q - \dfrac{\rho^2}{(q-\beta\Delta)^2}\Big( (q-\beta\Delta)\left( \rho^2\Delta q^{\prime\prime} - 2\rho^2q - 2r(q^\prime\Delta - q\Delta^\prime) \right) \nonumber\\
	&\hspace{1.3cm} + \rho^2(\kappa\rho^2 - q^\prime + \beta\Delta^\prime)(q^\prime\Delta - q\Delta^\prime) \Big) \bigg).
\end{align}
The different potentials for a given spin lead to the same results. In the potential for spin 2 particles, the following quantities appear
\begin{subequations}
	\begin{align}
		&q(r) \equiv \nu\rho^4 + 3\rho^2(r^2-a^2) - 3r^2\Delta\,, \\
		&q^\prime(r) = r\left( (4\nu + 6)\rho^2 - 6(r^2 - 3Mr + 2a^2) \right)\,, \\
		&q^{\prime\prime}(r) = (4\nu+6)\rho^2 + 8\nu r^2 - 6r^2 + 36Mr - 12a^2\,,
	\end{align}
\end{subequations}
\begin{equation}
	q^\prime\Delta - q\Delta^\prime = -2(r-M)\nu\rho^4 + 2\rho^2(2\nu r\Delta -3M(r^2 + a^2) + 6ra^2) + 12r\Delta(Mr - a^2)\,,
\end{equation}
\begin{equation}
	\beta_\pm \equiv \pm 3\alpha^2\,,
\end{equation}
\begin{equation}
	\kappa_\pm \equiv \pm\sqrt{36M^2 -2\nu(\alpha^2(5\nu+6)-12a^2) + 2\beta\nu(\nu+2)}\,,
\end{equation}
\begin{subequations}
	\begin{align}
		&q-\beta_+\Delta = \rho^2(\nu\rho^2 + 6Mr - 6a^2)\,, \\
		&q-\beta_-\Delta = \nu\rho^4 + 6r^2(\alpha^2-a^2) + 6Mr(r^2-\alpha^2)\,,
	\end{align}
\end{subequations}
where $\nu \equiv \lambda_{2\,lm} + 4$. Ref.~\cite{TorresdelCastillo:1992zq} provides ingredients for the spin $3/2$ which we have for now included only in the Schwarzschild case. The potential is in this case
\begin{align}
	V_{3/2}(r) = \dfrac{\Delta}{\rho^6}\Bigg( q &- \dfrac{1}{(q - \beta_\pm\sqrt{\Delta})^2}\Big( (q \sqrt{\Delta} - \beta_\pm\Delta)\left( \rho^2\sqrt{\Delta}q^{\prime\prime} - \rho^2 q\Delta^{\prime\prime} - 2r(q^\prime \sqrt{\Delta} - \Delta^\prime q) \right) \nonumber\\
	&+ \rho^2( \kappa_\pm\rho^2 - \sqrt{\Delta}q^\prime + q\Delta^\prime )(q^\prime\sqrt{\Delta} - \Delta^\prime q) \Big) \Bigg),
\end{align}
where the intermediate quantities are this time
\begin{equation}
	q(r) \equiv \lambda_{3/2,lm}\rho^2 + 2Mr - 2a^2\,,
\end{equation}
\begin{equation}
	\beta_\pm \equiv \pm \sqrt{4(a^2 - \lambda_{3/2,lm}\alpha^2)}\,,
\end{equation}
\begin{equation}
	\kappa_\pm \equiv \pm\sqrt{\lambda_{3/2,lm}^3 + \lambda_{3/2,lm}^2}\,.
\end{equation}

In the Schwarzschild limit ($a^* = 0$), we recover the Regge--Wheeler potentials~\cite{Regge:1957td}. As the angular momentum projection $m$ only appears multiplied by $a$, in this case the calculation is simplified since only one common value for all $m$ has to be chosen once $l$ is fixed. The sum over $m$ thus reduces to a factor $2l+1$ and the $r(r^*)$ relation of Eq.~\eqref{eq:tortoise} is analytical.

The $r^*$ variable change used in these potentials leads to divergences in the potentials, when $r^2 = r\mrm{div}^2 \equiv -\alpha^2$. This can happen for sufficiently low energies and high (negative) angular momentum projections, and it corresponds to the \textit{superradiance} regime (for a very good review on the topic, see~\cite{Brito:2015oca}). As discussed in the Chandrasekhar and Detweiler papers, the technique to avoid this divergence is to integrate Eq.~\eqref{eq:kerrRW} up to slightly before the divergence (\textit{i.e.} $r\mrm{div}-\epsilon$). At this point, the behaviour of the potential $V_s$ is known, Eq.~\eqref{eq:kerrRW} is simplified, and the asymptotic form of the function $\psi_s$ can be obtained for $\epsilon \rightarrow 0$. By continuity of the wave function $\psi_s$ one can extrapolate this form up to slightly \textit{after} the divergence (\textit{i.e.} $r\mrm{div}+\epsilon$) and continue the integration.

Another difficulty which can arise is the fact that there can be an additional divergence in the spin 2 or spin $3/2$ potentials because of the $q-\beta_\pm\Delta$ and $q - \beta_\pm\sqrt{\Delta}$ terms. For this extra divergence, we try to integrate with one of the potentials (\textit{e.g.} $\kappa_+$, $\beta_+$), and in case of a problem we try with the other potentials (\textit{e.g.} $\kappa_+$, $\beta_-$), as it seems that at least one of the four combinations does not generate any divergence.

\subsubsection{Spherically symmetric and static black holes}

For spherically symmetric and static BH solutions of general form
\begin{equation}
	\d s^2 = -G(r)\d t + \dfrac{1}{F(r)}\d r^2 + H(r)\d \Omega^2\,,
\end{equation}
the authors, among others, have recently shown how any of those BH metric leads to a separation between the angular part and the radial part, and finally to a Schr\"odinger-like wave equation of a form similar to Eq.~\eqref{eq:kerrRW} with the following potentials~\cite{Arbey:2021jif,Arbey:2021yke}
\begin{subequations}
	\begin{align}
		&V_0(r^*)=\nu_0\dfrac{G}{H}+\dfrac{\partial_*^2\sqrt{H}}{\sqrt{H}}\,,\\
		&V_1(r^*)=\nu_1\dfrac{G}{H}\,,\\
		&V_2(r^*)=\nu_2\dfrac{G}{H}+\dfrac{(\partial_*H)^2}{2H^2}-\dfrac{\partial_*^2\sqrt{H}}{\sqrt{H}}\,,\\
		&V_{1/2}(r^*)=\nu_{1/2}\dfrac{G}{H} \pm \sqrt{\nu_{1/2}}\,\partial_*\left( \sqrt{\dfrac{G}{H}} \right),
	\end{align}
\end{subequations}
where the spin-dependent parameters $\nu_i$ are given by $\nu_0 = l(l+1) = \nu_1$, $\nu_2 = l(l+1)-2$ and $\nu_{1/2} = l(l+1)+1/4$ and the Eddington--Finkelstein coordinate is defined by
\begin{equation}
	\dfrac{\d r^*}{\d r} \equiv \dfrac{1}{\sqrt{FG}}\,,
\end{equation}
with the notation $\partial_* \equiv \partial_{r^*}$. The $r(r^*)$ relation can be analytical or not, depending on the details of the metric (see Appendix A of~\cite{Arbey:2021yke}), but usually does not show a pathological behaviour. These potentials reduce to the Regge--Wheeler ones in the Schwarzschild limit $F(r) = G(r) = 1-r\mrm{S}/r$ and $H(r) = r^2$~\cite{Regge:1957td}.

\subsubsection{Boundary conditions}

As boundary conditions to solve Eq.~\eqref{eq:kerrRW}, we use a purely ingoing wave. The solution at the horizon has the form
\begin{equation}
	\psi_s = e^{iE r^*}\,. \label{eq:boundary1}
\end{equation}
At infinity, the solution has the form
\begin{equation}
	\psi_s = A\mrm{in}e^{iE r^*} + A\mrm{out}e^{-iE r^*}\,. \label{eq:boundary2}
\end{equation}
The greybody factor is given by the transmission coefficient of the wave from the horizon to the infinity
\begin{equation}
	\Gamma_{slm} \equiv T_{slm} = \dfrac{1}{|A\mrm{in}|^2}\,.
\end{equation}
Practically, we compute the value of a single helicity/color dof emissivity
\begin{equation}
	Q_s\equiv \displaystyle\sum_{l,m}\dfrac{\Gamma_{slm}}{\left(e^{E^\prime/T} - (-1)^{2s}\right)}\,,
\end{equation}
for:
\begin{itemize}
	\item 50 values of $0 \le a^* \le 0.9999$ for Kerr BHs,
	\item 50 values of $0\le Q^* \le 0.999$ for Reissner--Nordstr\"om BHs,
	\item 7 values of $0\le n \le 6$ for higher-dimensional BHs,
	\item 22 values of $0\le \varepsilon \le 100$ for polymerized BHs (with $a_0 = 0$ and $a_0 = 0.11$),
\end{itemize}
and for a range of 200 energies $0.01 \le x\equiv Er\mrm{S} \le 5$ (dimensionless). For $x$ out of this range, we have found easier to fit empiric asymptotic forms to the emissivities. At low energies, we have for all spins $s$
\begin{equation}
	\log_{10}(Q_s) \approx a_{1,s}\log_{10}(x) + a_{2,s}\,,
\end{equation}
and at high energies for the Kerr BH
\begin{equation}
	\log_{10}(Q_s) \approx a_{3,s} x + a_{4,s} + a_{5,s}\cos(a_{7,s} x) + a_{6,s}\sin(a_{7,s} x)\,,
\end{equation}
which allows for a reconstruction of the oscillatory behaviour in this limit~\cite{Dong:2015yjs}. For spherically symmetric and static BHs, we chose at high energy to keep only
\begin{equation}
	\log_{10}\left(Q_s\times(e^{E^\prime/T} - (-1)^{2s})/(27/4)E^2 M^2\right) \approx a_{8,s}\,,
\end{equation}
as the oscillatory behaviour is damped and the constant cross-section regime is attained rapidly~\cite{Arbey:2021yke}. We checked that the fitting parameters agree with the asymptotic limits of~\cite{MacGibbon:1990zk} in the Schwarzschild case and of~\cite{Page:1976df} in the Kerr case, and with analytical formulas we derived in~\cite{Arbey:2021yke} in the spherically-symmetric and static case. For Kerr BHs, the \texttt{Mathematica} scripts as well as a \texttt{C} formatting script and a \texttt{README} are provided in the folder:

\citecode{scripts/greybody\_scripts/greybody\_factors/spin\_*.m}

\citecode{scripts/greybody\_scripts/greybody\_factors/exploitation.m}

\citecode{scripts/greybody\_scripts/greybody\_factors/formatting.c}

\citecode{scripts/greybody\_scripts/greybody\_factors/Makefile}

\citecode{scripts/greybody\_scripts/greybody\_factors/README.txt}

\noindent For the other BH solutions, the \texttt{Mathematica} and \texttt{Python} scripts are in the relevant folders:

\citecode{scripts/greybody\_scripts/greybody\_factors/*/spin\_*.m}

\citecode{scripts/greybody\_scripts/greybody\_factors/*/exploitation\_*.py}

\noindent If these tables are recomputed, it is advisable to also modify the information in the file:

\citecode{src/tables/gamma\_tables/infos.txt}

\subsection{Page factors}
\label{app:fM_tables}

To compute the integrals of Eqs.~\eqref{eq:fMa} and \eqref{eq:gMa}, we use the greybody factor tables and the fits computed in the previous Section. The peak of Hawking emission lies around the BH temperature for spherically symmetric and static BHs (see \textit{e.g.}~\cite{Arbey:2021yke}), and is uncorrelated for Kerr BHs (see \textit{e.g.}~\cite{Dong:2015yjs}); in any case it is close to the Schwarzschild temperature for reasonable values of the secondary parameters. Thus the integral does not need to be computed over all energies, but a restrained set of energies in $10^{-5}\times T < E < 10^5\times T$ is sufficient. The domains of integration are segmented over 1000 logarithmically distributed energies, and computed for 1000 masses $M\mrm{P} < M < 10^{46}\,$GeV ($\sim 10^{-5} < M < 10^{22}\,$g). We compute these tables with and without the graviton emission, and with or without a single dof of DM emission. In the limit where neutrinos are massless, $f(M,x_j)$ and $g(M,a^*)$ are not expected to change for masses higher than $10^{22}\,$g, the tables can therefore be extended manually without any new computation. Masses are given in GeV (corresponding to grams) and $f(M,x_j)$ and $g(M,a^*)$ are in GeV$^4$ (corresponding to g$^3\cdot$s$^{-1}$ and g$^2\cdot$GeV$\cdot$s$^{-1}$, respectively). We have checked that the value of $f(M,0)$ is consistent with that of \cite{MacGibbon:1991tj} in the Schwarzschild case and that the values of $f(M,a^*)$ and $g(M,a^*)$ are consistent with \cite{Taylor:1998dk} in the Kerr case. No numerical results were ever given for the polymerized case. The C scripts used to compute the tables, a \texttt{README} and all the necessary greybody factors tables are provided in the subfolder:

\citecode{scripts/greybody\_scripts/fM/fM.c}

\citecode{scripts/greybody\_scripts/fM/Makefile}

\citecode{scripts/greybody\_scripts/fM/README.txt}

\citecode{scripts/greybody\_scripts/fM/spin\_*.txt}

\noindent If these tables are recomputed, it is advisable to also modify the information in the file:

\citecode{src/tables/fM\_tables/infos.txt}

\subsection{Hadronization tables}
\label{app:hadronization_tables}

Three particle physics codes have been used to compute hadronization tables: \texttt{PYTHIA}~\cite{Sjostrand:2014zea}, \texttt{HERWIG}~\cite{Bellm:2015jjp} and \texttt{Hazma}~\cite{Coogan:2019qpu}. In the first two cases, the strategy is to generate the output of a collision (for example $e^+ + e^-\rightarrow u + \overline{u}\rightarrow \dots$), and then to count the number of final particles (here denoted as dots) normalized by the number (here 2) of initial particles (here $u$) satisfying the desired stability criterion: Table~\ref{tab:BBN_particles} for early Universe/BBN particles (\texttt{pythia\_tables} and \texttt{herwig\_tables}) and Table~\ref{tab:today_particles} for present epoch particles (\texttt{pythia\_tables\_new}). This strategy is adapted for those two codes which are Monte-Carlo simulators. \texttt{Hazma} is quite different as it computes analytically the branching ratios once the initial energy is fixed.

To build the \texttt{PYTHIA} and \texttt{HERWIG} tables, we have simulated for each channel listed in Table~\ref{tab:hadro}, $10^5$ events for initial energies $E^\prime$ (half of the center of mass energy) logarithmically distributed between $5\,$GeV and $10^5\,$GeV (\texttt{PYTHIA}) or between $25\,$GeV and $10^5\,$GeV (\texttt{HERWIG}). Then, the final particles have been listed as a function of their final energy $E$, into a range of $10^{-6}\,$GeV to $10^5\,$GeV and the counts have been averaged over the number or simulated events. This gives the dimensionless quantities ${\rm d}N_j^i(E^\prime,E)$ of Eq.~\eqref{eq:secondary}. For the \texttt{Hazma} tables, we have directly applied the analytical computation of the branching ratios to arrays of initial energies $10^{-6}-5\,$GeV for all the available particles: electrons, muons and pions. The procedure is well described in~\cite{Coogan:2019qpu,Coogan:2020tuf}.

\begin{table}[!ht]
	\centering{
		\begin{tabular}{|c|c|c|}
			\hline
			particle & \texttt{PYTHIA} (new) & \texttt{HERWIG} \\
			\hline
			gluons & $e^+ e^-\,\rightarrow\, h^0\,\rightarrow\,g\overline{g}$ & $e^+ e^-\,\rightarrow\, h^0\,\rightarrow g\overline{g}$ \\
			Higgs boson	& $e^+ e^-\,\rightarrow\, h^0$ & $e^+ e^-\,\rightarrow\, h^0$ \\
			W bosons & $e^+ e^-\,\rightarrow\,Z^0/\gamma^*\,\rightarrow\,W^+ W^-$ & $e^+ e^-\,\rightarrow\,Z^0/\gamma^*\,\rightarrow\,W^+ W^-$ \\
			Z boson	& $e^+ e^-\,\rightarrow\, h^0\,\rightarrow\,Z^0 Z^0$ & $e^+ e^-\,\rightarrow\,Z^0/\gamma^*\,\rightarrow\,Z^0 Z^0$ \\
			leptons	& $e^+ e^-\,\rightarrow\, h^0\,\rightarrow\,l^+ l^-$ & $e^+ e^-\,\rightarrow\,Z^0/\gamma^*\,\rightarrow\,l^+ l^-$ \\
			quarks & $e^+ e^-\,\rightarrow\,Z^0/\gamma^*\,\rightarrow\,q\overline{q}$ & $e^+ e^-\,\rightarrow\,Z^0/\gamma^*\,\rightarrow\,q\overline{q}$ \\
			\hline
		\end{tabular}
		\caption{List of the channels used to compute the hadronization tables for the Monte-Carlo codes.\label{tab:hadro}}
	}
\end{table}

The branching ratios $e^\pm \rightarrow \gamma\gamma \rightarrow ...$ and $e^\pm \rightarrow \nu\overline{\nu}\rightarrow ...$ have not been computed with \texttt{PYTHIA} and \texttt{HERWIG}. The contribution from the primary photons, neutrinos and electrons are directly added to the secondary spectra with a branching ratio of 1, which is an approximation. \texttt{Hazma}, on the other hand, handles the final state radiation and the decay of pions and muons into photons and electrons, thus in this case only the neutrino final spectrum is out of reach.

For initial energies lower than the cutoff of the \texttt{PYTHIA} and \texttt{HERWIG} tables, branching ratios from the lowest relevant initial energy will be extrapolated at lower energies once shifted to the considered energy, taking into account that no emission can arise below the rest mass of the final particles. This is a mere numerical trick that was shown by~\cite{release_note,Coogan:2020tuf} to be irrelevant for BH masses close to the QCD scale $M\sim 10^{15}\,$g. To be clear: \texttt{PYHTIA} and \texttt{HERWIG} are particle physics codes that are very efficient in their domain of validity, for energies between some GeV and some TeV (BH mass $M < 10^{13}\,$g), and considers SM particles (\textit{e.g.}~quarks) as elementary degrees of freedom. However, \texttt{Hazma} is much more fitted to compute the low energy showering of (charged) particles and considers pions as fundamental degrees of freedom below the QCD scale of $\sim 200\,$MeV (BH mass $M \gtrsim 10^{14}\,$g). As their is no sizeable secondary generated particles for energies below the MeV scale (BH mass $M \gtrsim 10^{17}\,$g), all these codes should provide the same result.

The scripts used to compute the \texttt{pythia\_tables} are given in the folder:

\citecode{scripts/pythia\_scripts/formatting.c}

\citecode{scripts/pythia\_scripts/Makefile}

\citecode{scripts/pythia\_scripts/README.txt}

\citecode{scripts/table\_*.cc}

\noindent The scripts used to compute the \texttt{pythia\_tables\_new} are given in the folder:

\citecode{scripts/pythia\_new\_scripts/formatting.c}

\citecode{scripts/pythia\_new\_scripts/Makefile}

\citecode{scripts/pythia\_new\_scripts/README.txt}

\citecode{scripts/pythia\_new\_scripts/table\_*.cc}

\noindent The scripts used to compute the \texttt{herwig\_tables} are given in the folder:

\citecode{scripts/herwig\_scripts/formatting.c}

\citecode{scripts/herwig\_scripts/Makefile}

\citecode{scripts/herwig\_scripts/README.txt}

\citecode{scripts/herwig\_scripts/BH\_TABLE\_GENERATOR.cc}

\citecode{scripts/herwig\_scripts/MC\_PARTICLE\_COUNTS.cc}

\citecode{scripts/herwig\_scripts/RivetMC\_PARTICLE\_COUNTS.so}

\citecode{scripts/herwig\_scripts/*/LEP.in}

\citecode{scripts/herwig\_scripts/*/main.cpp}

\noindent The script used to compute the \texttt{Hazma} tables is given in the folder:

\citecode{scripts/hazma\_scripts/hazma\_tables.py}

\noindent Please contact one of the authors if you have issues using these scripts. If these tables are recomputed, it is advisable to also modify the informations in the files:

\citecode{src/tables/hadronization\_tables/infos.txt}

\section{Example runs}
\label{app:example_results}

In this Section, we provide the expected results of a run of each of the \texttt{BlackHawk} programs, when using the built-in \texttt{parameters.txt} file. They should be strictly equivalent to the user's results.

\subsection{Built-in parameters.txt file}
\label{app:example_parameters}

The built-in parameters are:\newline
\texttt{
	\noindent destination\_folder = example\newline
	full\_output = 1\newline
	interpolation\_method = 0\newline \newline
	BH\_number = 1\newline
	Mmin = 1e+9\newline
	Mmax = 1e+16\newline
	metric = 0\newline
	param\_number = 1\newline
	amin = 0.\newline
	amax = 0.5\newline
	Qmin = 0.\newline
	Qmax = 0.7\newline
	epsilon\_LQG = 0.\newline
	a0\_LQG = 0.\newline
	n = 0.\newline \newline
	spectrum\_choice = 0\newline
	spectrum\_choice\_param = 0\newline \newline
	amplitude\_lognormal = 1.\newline
	amplitude\_lognormal2 = 1.\newline
	stand\_dev\_lognormal = 1.\newline
	crit\_mass\_lognormal = 1.\newline \newline
	amplitude\_powerlaw = 1.\newline
	eqstate\_powerlaw = 0.3333\newline \newline
	amplitude\_critical\_collapse = 1.\newline
	crit\_mass\_critical\_collapse = 1.\newline \newline
	amplitude\_uniform = 1.\newline \newline
	stand\_dev\_param\_gaussian = 1.\newline
	mean\_param\_gaussian = 0.5\newline \newline
	table = spin\_distribution\_BH.txt\newline \newline
	tmin\_manual = 0\newline
	tmin = 1.e-30\newline
	limit = 5000\newline
	BH\_remnant = 0\newline
	M\_remnant = 1e-4\newline \newline
	E\_number = 10\newline
	Emin = 5\newline
	Emax = 1e+5\newline \newline
	grav = 0\newline
	add\_DM = 0\newline
	m\_DM = 0.\newline
	spin\_DM = 0.\newline
	dof\_DM = 0.\newline \newline
	primary\_only = 0\newline \newline
	hadronization\_choice = 0\newline
}

\subsection{BlackHawk\_tot output files}

When launching \texttt{BlackHawk\_tot} with the built-in \texttt{parameters.txt} file, one obtains the following output files (truncated to 5 lines and 5 rows when relevant):
\begin{itemize}
	\item \texttt{BH\_spectrum.txt}
	\begin{flushleft}
		\texttt{
			Initial BH comoving number density as a function of their mass and parameter.\newline
			\begin{tabular}{rr}
				mass/spin &   0.00000e+00\\
				1.00000e+09 &   1.00000e+00\\
			\end{tabular}
		}
	\end{flushleft}
	
	\item \texttt{life\_evolutions.txt}
	\begin{flushleft}
		\texttt{
			Evolution of the BH masses and spins as functions of time.\newline
			Total number of time iterations: 1044\newline
			\begin{tabular}{rrr}
				t&                   M&              a \\
				1.23833e-29    &     1.00000e+09   & 0.00000e+00\\
				2.47665e-29    &     1.00000e+09   & 0.00000e+00\\
				3.71498e-29    &     1.00000e+09   & 0.00000e+00\\
				4.95331e-29    &     1.00000e+09   & 0.00000e+00\\
				6.19164e-29    &     1.00000e+09   & 0.00000e+00
			\end{tabular}
		}
	\end{flushleft}
	
	\item \texttt{dts.txt}
	\begin{flushleft}
		\texttt{
			Evolution of the integration timestep as a function of time.\newline
			\begin{tabular}{rr}
				t           &  dt\\
				1.23833e-29  &  1.23833e-29\\
				2.47665e-29  &  1.23833e-29\\
				3.71498e-29  &  1.23833e-29\\
				4.95331e-29  &  1.23833e-29\\
				6.19164e-29  &  1.23833e-29
			\end{tabular}
		}
	\end{flushleft}
	
	\item \texttt{photon\_primary\_spectrum.txt}
	\begin{flushleft}
		\texttt{
			Hawking primary spectrum as a function of time.\newline
			\begin{tabular}{rrrrrr}
				time/energy  &  5.00000e+00   & 1.50267e+01  &  4.51601e+01   & 1.35721e+02 & 4.07886e+02 \\
				1.23833e-29 &   4.10141e+09 &   1.02586e+11 &   2.56592e+12 &   6.41798e+13 & 1.60529e+15\\
				2.47665e-29 &   4.10141e+09 &   1.02586e+11 &   2.56592e+12  &  6.41798e+13 & 1.60529e+15\\
				3.71498e-29  &  4.10141e+09 &   1.02586e+11 &   2.56592e+12   & 6.41798e+13 & 1.60529e+15\\
				4.95331e-29  &  4.10141e+09 &   1.02586e+11 &   2.56592e+12    &6.41798e+13 & 1.60529e+15\\
				6.19164e-29  &  4.10141e+09 &   1.02586e+11 &   2.56592e+12   & 6.41798e+13   & 1.60529e+15
			\end{tabular}	
		}
	\end{flushleft}
	
	\item \texttt{photon\_secondary\_spectrum.txt}
	\begin{flushleft}
		\texttt{
			Hawking secondary spectrum as a function of time.\newline
			\begin{tabular}{rrrrrr}
				time/energy  &  1.00000e-06   & 1.05207e-06  &  1.10685e-06 &   1.16448e-06  &  1.22511e-06\\
				1.23833e-29   & 9.55768e+30   & 1.14321e+31  &  1.87205e+31  &  1.90536e+31   & 1.92737e+31\\
				2.47665e-29  &  9.55768e+30   & 1.14321e+31  &  1.87205e+31   & 1.90536e+31   & 1.92737e+31\\
				3.71498e-29 &   9.55768e+30   & 1.14321e+31  &  1.87205e+31   & 1.90536e+31   & 1.92737e+31\\
				4.95331e-29  &  9.55768e+30   & 1.14321e+31  &  1.87205e+31  &  1.90536e+31   & 1.92737e+31\\
				6.19164e-29  &  9.55768e+30   & 1.14321e+31   & 1.87205e+31   & 1.90536e+31   & 1.92737e+31
			\end{tabular}	
		}
	\end{flushleft}
\end{itemize}

\subsection{BlackHawk\_inst output files}

When launching \texttt{BlackHawk\_inst} with the built-in \texttt{parameters.txt} file, one obtains the following output files (truncated to 5 lines and 5 rows when relevant):
\begin{itemize}
	\item \texttt{BH\_spectrum.txt} is the same as for \texttt{BlackHawk\_tot}
	
	\item \texttt{instantaneous\_primary\_spectra.txt}
	\begin{flushleft}
		\texttt{
			Hawking primary spectra for each particle types.\newline
			\begin{tabular}{rrrrrr}
				energy/particle   &      photon   &      gluons  &        higgs       &     W+-     &        Z0\\
				5.00000e+00   & 4.10141e+09   & 3.28113e+10   & 0.00000e+00   & 0.00000e+00   & 0.00000e+00\\
				1.50267e+01   & 1.02586e+11   & 8.20688e+11  &  0.00000e+00   & 0.00000e+00   & 0.00000e+00\\
				4.51601e+01   & 2.56592e+12   & 2.05274e+13  &  0.00000e+00   & 0.00000e+00   & 0.00000e+00\\
				1.35721e+02   & 6.41798e+13   & 5.13438e+14   & 8.96189e+19   & 1.92539e+14   & 9.62696e+13\\
				4.07886e+02   & 1.60529e+15   & 1.28423e+16   & 2.50679e+20   & 4.81586e+15   & 2.40793e+15
			\end{tabular}	
		}
	\end{flushleft}
	
	\item \texttt{instantaneous\_secondary\_spectra.txt}
	\begin{flushleft}
		\texttt{
			Hawking secondary spectra for each particle types.\newline
			\begin{tabular}{rrrrrr}
				energy/particle  &       photon    &   electron    &       muon    &       nu\_e  &        nu\_mu\\
				1.00000e-06   & 9.55768e+30   & 0.00000e+00 &   0.00000e+00   & 0.00000e+00   & 0.00000e+00\\
				1.05207e-06   & 1.14321e+31  &  0.00000e+00   & 0.00000e+00 &    0.00000e+00   & 0.00000e+00\\
				1.10685e-06   & 1.87205e+31   & 0.00000e+00   & 0.00000e+00 &   0.00000e+00   & 0.00000e+00\\
				1.16448e-06   & 1.90536e+31   & 0.00000e+00   & 0.00000e+00  &  0.00000e+00   & 0.00000e+00\\
				1.22511e-06   & 1.92737e+31   & 0.00000e+00   & 0.00000e+00   & 0.00000e+00   & 0.00000e+00
			\end{tabular}	
		}
	\end{flushleft}
\end{itemize}

\bibliography{biblio}

\end{document}